\documentclass[epj]{svjour}
\usepackage{graphics}
\usepackage[dvips]{epsfig}
\usepackage{graphicx} 
\usepackage{lscape}
\usepackage{multirow}
\usepackage{lineno}
\begin{document}
%

\newcommand{\sip}{AgReO$_{4}$}
\newcommand{\ti}{$^{44}$Ti}
\newcommand{\ka}{K$\alpha$}
\newcommand{\kaone}{K$\alpha_1$}
\newcommand{\katwo}{K$\alpha_2$}
\newcommand{\kb}{K$\beta$}
\newcommand{\rhot}{$R_{hot}$}
\def\Journal#1#2#3#4{{#1} {\bf #2}, #4 (#3)} 
\def\NCA{Nuovo Cimento}
\def\NIM{Nucl.\,Instrum.\,Methods}
\def\NIMA{Nucl.\,Instrum.\,Methods A}
\def\NIMB{Nucl.\,Instrum.\,Methods B}
\def\NPB{Nucl.\,Phys. B}
\def\NP{Nucl.\,Phys.}
\def\NPBP{Nucl.\,Phys. B (Proc. Suppl.)}
\def\PLB{Phys.\,Lett.  B}
\def\PL{Phys.\,Lett.}
\def\JPD{J.\,Phys. D}
\def\JPG{J.\,Phys. G}
\def\PRL{Phys.\,Rev.\,Lett.}
\def\PRD{Phys.\,Rev. D}
\def\PRC{Phys.\,Rev. C}
\def\PR{Phys.\,Rev.}
\def\ZPC{Z.\,Phys. C}
\def\EPL{Europhys.\,Lett.}
\def\JP{J.\,Phys.}
\def\AP{Astropart.\,Phys.}
\def\RP{Phys.\,Rep.}
\def\ARNPS{Annu.\,Rev.\,Nucl.\,Part.\,Sci.}
\def\JLT{J.\,Low\,Temp.\,Phys.}

\def\bbn{$\beta\beta$-0$\nu$}
\def\Te{$^{130}$Te}
\def\Fe{$^{55}$Fe}
\def\Ti{$^{44}$Ti}
\def\Re{$^{187}$Re}
\def\b{$\beta$}
\def\g{$\gamma$}
\def\agre{AgReO$_4$}
\def\teo{TeO$_2$}
\def\Tr{$^3$H}
\def\mn{$m_\nu$}
\def\mne{$m_{\overline{\nu}_e}$}
\def\mnesq{$m_{\overline{\nu}_e}^2$}
\def\mug{$\mu$g}
\def\mum{$\mu$m}
\def\mumsq{$\mu$m$^2$}
\def\mumq{$\mu$m$^3$}
\def\mus{$\mu$s}
\def\halft{$\tau_{1/2}$}
\def\de{$\Delta E$}
\def\fwhm{$_{\mathrm{FWHM}}$}

\title{Investigation of peak shapes in the MIBETA experiment calibrations}

\author{E.\,Ferri\inst{1,2}, S.\,Kraft-Bermuth\inst{3}\thanks{corresponding
    author, email: saskia.kraft-bermuth@iamp.physik.uni-giessen.de},
A.\,Monfardini\inst{4}, A.\,Nucciotti\inst{1,2}, D.\,Schaeffer\inst{5}, M.\,Sisti\inst{1,2}
}                     
%
%
\institute{Dipartimento di Fisica dell'Universit\`a di Milano-Bicocca, 20126 Milano, Italia
\and INFN Sezione di Milano-Bicocca, 20126 Milano, Italia
\and Institut f\"ur Atom- und Molek\"ulphysik, Justus-Liebig-Universit\"at Gie\ss en, 35392 Gie\ss en, Deutschland
\and Institut N\'eel, CNRS \& Universit\'e Joseph Fourier (UJF), BP 166, 38042 Grenoble, France
\and ABB AB, Corporate Research, 72178 V\"aster\aa{}s , Sweden
}
\authorrunning{E.\,Ferri et al.}
\titlerunning{Eur. Phys. J. A (2012) {\bf 48}:131}

\date{Received: 5 April 2012 / Revised: 13 August 2012 \\ Published online: 12
October 2012}
%
\abstract{
In calorimetric neutrino mass experiments, where the shape of a beta decay spectrum has to be precisely measured, the understanding of the detector response function is a fundamental issue. In the MIBETA neutrino mass experiment, the X-ray lines measured with external sources did not have Gaussian shapes, but exhibited a pronounced shoulder towards lower energies. If this shoulder were a general feature of the detector response function, it would distort the beta decay spectrum and thus mimic a non-zero neutrino mass. An investigation was performed to understand the origin of the shoulder and its potential influence on the beta spectrum. First, the peaks were fitted with an analytic function in order to determine quantitatively the amount of events contributing to the shoulder, also depending on the energy of the calibration X-rays. In a second step,  Montecarlo simulations were performed to reproduce the experimental spectrum and to understand the origin of its shape. 
We conclude that at least part of the observed shoulder can be attributed to a surface effect. 
} 
\maketitle
\section{Introduction}
Low temperature microcalorimeters are expected to play a key role in future direct neutrino mass measurements \cite{enss-cryo}. With these detectors it is possible to perform very sensitive beta end-point studies in a calorimetric configuration, i.e. with the beta source embedded in the detector. In this way, the detector measures all the energy released in the beta decay except that carried away by the neutrino, and many systematic effects showing up in other configurations are avoided.
Nevertheless past neutrino mass experiments with low temperature microcalorimeters have shown how critical is the understanding of 
the detector response function.

In past years few calorimetric experiments using \Re\ as beta decaying isotope have been carried out both in Milano (MIBETA) and in Genova (MANU). A final
sensitivity on the neutrino mass of around 15\,eV was achieved by both groups \cite{MIBETA-PRL,MIBETA,MANU,MANU-PRC}.
In these experiments the neutrino mass squared $m_\nu^2$ is measured by looking at the tiny deficit of events in an energy interval below the beta decay end-point as small as few times the neutrino mass itself. For this purpose, the experimentally measured beta decay spectrum has to be interpolated by the function obtained as convolution of the theoretical beta decay spectrum with the detector response function $F(E,E_0)$ -- defined as the measured response for the deposition of a fixed amount of energy $E_0$. A precise evaluation of $F(E,E_0)$ is therefore mandatory to avoid systematic uncertainties.
The response function $F(E,E_0)$ is expected to be a symmetric Gaussian line centered at $E_0$ whose width is caused by the electronic noise and is independent of
$E_0$. In practice, however, the width of the response function depends on $E_0$ and is larger than expected assuming only the electronic noise contribution. Moreover, often the line is not symmetric and, even worse, its shape may depend on where $E_0$ is deposited inside the detector. 
Usually the response function $F(E,E_0)$ is investigated by means of suitable calibration sources, but this may be a source of systematic effects. First of all, it is almost impossible to have really mono-energetic sources: the understanding of the calibration source emission spectrum is therefore part of the game. Second, it is very difficult to emulate with a calibration source the beta decay energy deposition: the calibration source has to cover the same energy range as the beta particles, the energy depositing interactions must be spatially equivalent both in their global distribution throughout the detector and in the single tracks, and the energy deposition must be done by the same type of particle, i.e. an electron. Last but not least, it must be possible to remove the calibration source during the beta decay measurement without affecting the detector. 

In calorimetric neutrino mass experiments, given the low energy end-point of the beta decays (e.g. \Re\ end-point is 2.465\,keV \cite{MIBETA}), the only viable approach is to use an external low-energy X-ray source\footnote{The use of an external low energy monochromatic beta source would be much more problematic as shown by spectrometric neutrino mass experiments.}. The energy depositions are then caused mostly by the primary photo-electrons and by the cascade of secondary X-rays and Auger electrons. 
Few differences to the beta decay electron interactions are immediately apparent. First, whereas the creation of a photo-electron by an X-ray is followed by a cascade of secondary atomic radiation (mostly Auger electrons), low energy betas produce very little secondary radiation (mostly photons).
Second, with the X-rays it is experimentally difficult to achieve a uniform illumination of the active detector volume as it happens with the beta particles. It is therefore extremely important to understand the measured response function $F(E,E_0)$ in order to disentangle the contributions to its shape caused by the X-rays from an external source.

The first aim of the present work is to obtain a precise functional description of the detector response function $F(E,E_0)$ measured
for X-ray absorption. Then we try to establish whether and how this response
function can be used to analyze the beta decay spectrum. 
To accomplish this, the physical processes underlying the response function for X-ray absorption will be investigated comparing different hypotheses with the experimental
data. Furthermore, processes taking place in the source assembly that can cause low energy tails have to be carefully investigated.
Once a model is confirmed, Montecarlo simulations can be used to indirectly obtain the energy response for the 
beta decay electrons.

\section{Experimental Set-up and Data}
The present work is based on the data collected during the activity of the MIBETA experiment from 2000 to 2004 \cite{MIBETA-PRL,MIBETA}.
In particular,  the three runs named RUN9, RUN14 and RUN15 are considered. 
These three runs were carried out with the same detectors but with different configurations because of their different goals (see Table~\ref{tab:runs} for details on the set-ups).   
The detectors are ten microcalorimeters made of small \agre\ crystals glued to doped silicon chip thermometers. 
The crystal mass ranges from 250 to 300 \mug, for a total mass of about 2.68\,mg (see \cite{MIBETA} for details on the set-up).
The major differences between the three runs are their durations and the calibration source assemblies, although they share the same low energy X-ray calibration source based on the fluorescence of low Z materials exposed to the X-rays of two $^{55}$Fe sources. 

\begin{table}
\begin{tabular}{cccccc} 
\hline
RUN & Calibration & \multicolumn{3}{c}{Calibration sources} & Usable \\
ID & time [h] & $^{55}$Fe & Pb shield & $^{44}$Ti &  detectors \\
\hline
RUN9 & 240 & yes & no & no  & 8 \\
RUN14 & 1000 & yes & yes & no  & 8 \\
RUN15 & 1000 & yes & yes & yes  & 2 \\
\hline
\end{tabular}
\caption{\label{tab:runs} Summary of the experimental conditions in the three runs considered in this work.
The $^{55}$Fe calibration source refers to
the fluorescence source described in the text. The last column reports the
number of detectors which were used for the analysis presented in this work (see text).}
\end{table}

RUN9 was intended to be the MIBETA high statistics measurement with the goal of achieving the best statistical sensitivity on the neutrino mass and it features
a first unshielded version of the fluorescence calibration source. The measurement was stopped after only 1000 hours because of the high continuum background caused by the $^{55}$Fe inner Bremsstrahlung. 
In RUN14 a lead shielding for the $^{55}$Fe sources brought a drastic reduction of the background and the high statistics measurement was completed as planned.
RUN15 was carried out with an additional $^{44}$Ti source with the aim of investigating the detector energy response.

\begin{figure}
\resizebox{.45\textwidth}{!}{ 
\includegraphics*[]{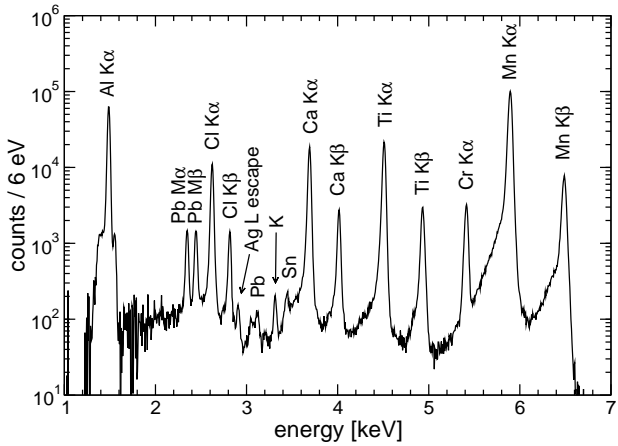}}
\caption{\label{fig:measured} Summed calibration spectrum of eight detectors from the RUN14 measurement. The  beta spectrum has been subtracted (see text). }
\end{figure}

The fluorescence source is made of two primary 5 mCi $^{55}$Fe sources irradiating two composite targets containing Al, CaF$_{2}$, Ti, and NaCl. Therefore, the detectors are exposed to the K fluorescence lines of Al, Ca, Ti and Cl as well as to the Rayleigh scattered K X-rays of Mn resulting from the electron capture decay of $^{55}$Fe. 
Thanks to a mechanism operated from outside the cryostat, the  $^{55}$Fe sources can be shut off in order to stop the emission of the fluorescent X-rays and to measure the beta decay spectrum without background.  
In RUN9 the $^{55}$Fe sources are shut off by two copper foils, while in RUN14 and RUN15 the $^{55}$Fe sources are moved inside a thick lead shielding.
In all runs, additional weak fluorescence lines are observed because of the
several materials present in the set-up, like the Pb used for the primary
source housing in RUN14 and RUN15 or the stainless steel used for the primary source rails and case. In particular, in the RUN14 spectrum -- shown in 
Figure~\ref{fig:measured} -- the Pb M lines below 2.5\,keV  and the Cr \ka\  line at 5.415 keV are evident.
Table~\ref{tab:energies1} lists the relevant X-ray energies together with their respective ranges in \sip. 

\begin{table}[b]
\begin{center}

\begin{tabular}{l@{\hspace{1cm}}cc}\hline
line & energy &  atten. length\\
 & [keV] & [$\mu$m]\\ \hline
\ti\ $\gamma$ & 78.337 & 268.3\\
\hline
Mn \kb\  & 6.490  & 5.3  \\
Mn \ka\  & 5.899  & 4.1  \\
Cr \ka\  & 5.415  & 3.3  \\
Ti \kb\  & 4.932  & 2.6  \\
Ti \ka\  & 4.511  & 2.1  \\
Ca \kb\  & 4.014  & 1.6  \\
Ca \ka\  & 3.691  & 1.3  \\
Cl \kb\  & 2.818  & 1.0  \\
Cl \ka\  & 2.622  & 0.9  \\
Pb M$\beta$ & 2.444 & 0.8 \\
Pb M$\alpha$ & 2.347 &  0.8  \\
Al \ka\  & 1.486  & 0.7 \\ 
\hline
\end{tabular}
\caption{\label{tab:energies1} Energies \cite{zsch07} and ranges of calibration $\gamma$ and X-rays (only \kaone\ and \kb\ are listed). 
} 
\end{center}

\begin{center}
\begin{tabular}{lcccc}\hline
&  peak position& \multicolumn{3}{c}{Rhenium X-rays}\\
line name & & energy & $\mu$ & range\\
 &[keV]& [keV] & [$\mu$m$^{-1}$] & [$\mu$m]\\
\hline
Re \ka\  1 escape & 17.196 &61.141 & 0.044 & 23 \\
Re \ka\  2 escape & 18.618 & 59.717& 0.035 & 28 \\ 
Re \kb\  1 escape & 9.027 & 69.31& 0.024 & 42 \\ 
\hline
\end{tabular}
\caption{\label{tab:energies2} Escape peaks: positions together with the energies \cite{zsch07} and ranges of X-rays involved in the escape process.}
\end{center}
\end{table}

\begin{figure}
\resizebox{.45\textwidth}{!}{ 
\includegraphics*[]{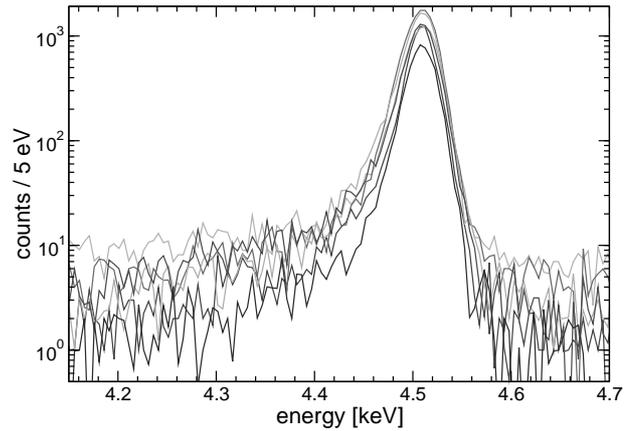}}
\caption{\label{fig:singlechannelTi} Ti \ka\ lines from the calibration spectrum of each channel which is included in the sum of Figure\,\ref{fig:measured}. In spite of the lower statistics all spectra show the low energy tail observed in the summed spectrum.}
\end{figure}

The data of RUN14 used for the present work are the subset collected during the calibration cycle of the measurements with only eight of the ten detectors. The total measurement time amounts to about 1000 hours and the resulting spectrum is shown in 
Figure~\ref{fig:measured}. The beta spectrum of \Re\ has been subtracted  -- about $6\times10^5$\,counts between 1.0 and 2.465\,keV--
to enhance the  Al lines and ease their investigation. 
Unfortunately their shape has been altered by the software cuts applied to the data in order to separate spurious events from true beta decays (see \cite{MIBETA} for details about the data analysis). 
It can be clearly seen in Figure~\ref{fig:measured} that all lines deviate from 
Gaussian  shapes, clearly exhibiting a tail towards lower energies. In spite of the lower statistics, Figure\,\ref{fig:singlechannelTi} shows that also the lines in the individual channel spectra have the same low energy tail. The FWHM
energy resolution on the Cl K$_\alpha$ peak -- the closest to the beta decay end-point -- is about 29\,eV. The same tails are actually observed every time calibration spectra with high statistics are collected with \agre\ detectors.

As a first step to understand the observed calibration spectrum, two Montecarlo simulations were performed using the Geant4 toolkit \cite{Geant4}.
The first simulates simply the direct interaction of
mono-energetic X-rays in the \agre\ absorbers showing that no tails are
expected in the fluorescence peaks due to mechanisms like emission and loss of
secondary radiation following the photoelectric absorption. 

\begin{figure}
\begin{center}
\resizebox{.45\textwidth}{!}{ 
\includegraphics*[]{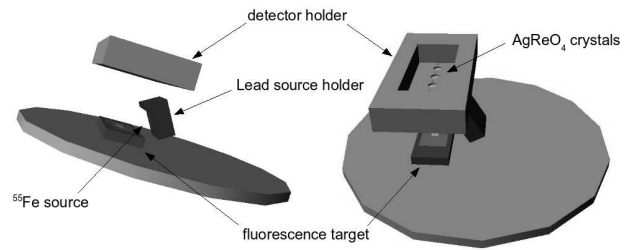}}
\caption{\label{fig:setupMC} Virtualized representation of the
  experimental setup with the $^{55}$Fe source and the fluorescence target.}
\end{center}
\end{figure}

\begin{figure}
\resizebox{.45\textwidth}{!}{ 
\includegraphics*[]{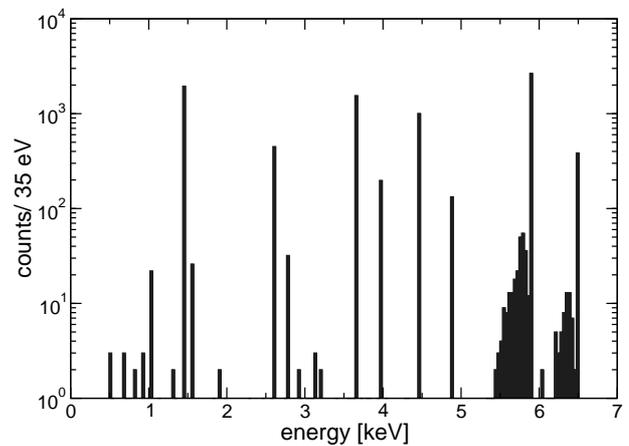}}
\caption{\label{fig:mc_nuovo} This simulation of the $^{55}$Fe fluorescence
  spectrum shows that no shoulders are expected for the fluorescence
  peaks. The shoulders in the Mn peaks stem from Compton back-scattering.}
\end{figure}

The second is a detailed Montecarlo simulation of the whole experimental set-up. 
Figure\,\ref{fig:setupMC} shows a virtualized representation of the simulated set-up. The spectrum in Figure\,\ref{fig:mc_nuovo} was obtained using the Penelope low energy extension of Geant4 \cite{Geant4}. The standard Geant4 low energy
extension was discarded since it introduced in the spectrum extra features which were not seen in the experimental data\footnote{In order to obtain correct results from Geant4, we have modified the Livermore Evaluated Atom Data Library (EADL) fluorescence files for the 
atomic number of interest because the original one had wrong X-ray transition energies.}. The Montecarlo spectrum corresponds to about $2\times10^{11}$ generated $^{55}$Fe decays -- i.e. to about one hour of experimental live time -- including all
X-rays and all Auger electrons. 
The spectrum in Figure\,\ref{fig:mc_nuovo} clearly shows that no tails are expected in the fluorescence peaks due to scattering in the set-up. 
At the same time the simulation shows that a large exponential tail is expected for the Mn Rayleigh scattered peaks due to a small fraction
of Compton back-scattering events. The exponential tail contains about 10\% of the total Mn \ka\ events and has a decay constant $\lambda$ of about 10\,keV$^{-1}$.

In order to understand the origin of the tail observed in all other peaks, as mentioned above, it is important to take into account the different spatial distributions of X-ray and \Re\ decay events. 
As can be seen from Table\,\ref{tab:energies1}, the calibration X-rays have ranges in \sip\ of 1--5 $\mu$m only. As the \sip\ crystals have an average thickness of  240~$\mu$m, this means that the energy deposition of the X-rays happens only in a thin surface layer containing  less than 2\% of the total crystal volume, while the beta decays occur uniformly across the whole crystal volume. 
The purpose of RUN15 was to experimentally investigate whether such a tail
occurs also for events depositing energy inside the crystal volume like the \Re\
beta decays. To accomplish this, a $^{44}$Ti source was introduced into the
set-up. $^{44}$Ti decays by electron capture to $^{44}$Sc emitting also one
$\gamma$-ray with an energy of about 78.337\,keV and a branching ratio of
about 98\%. Its energy is therefore just above the K-edge of Rhenium at
71.7\,keV, so peaks at about 18, 17 and 9\,keV due to the escape of \kaone,
\katwo\ and \kb\ Re X-rays, respectively, are expected in the experimental
spectrum. 
Since for an energy of about 78\,keV the attenuation length in \agre\ is larger than 200\,\mum\ the escape peaks, in spite of being located in the low energy region of the spectrum, are the results of interactions distributed almost uniformly in the crystal.
The energies and ranges related to the escape process are listed in Table\,\ref{tab:energies2}. 
To excite the escape peaks a $\gamma$-ray is much preferred over a X-ray in order
to limit the peak broadening to just the contribution of the Re atomic transition natural width.
\begin{figure}
\begin{center}
\resizebox{.45\textwidth}{!}{ 
 \includegraphics*[]{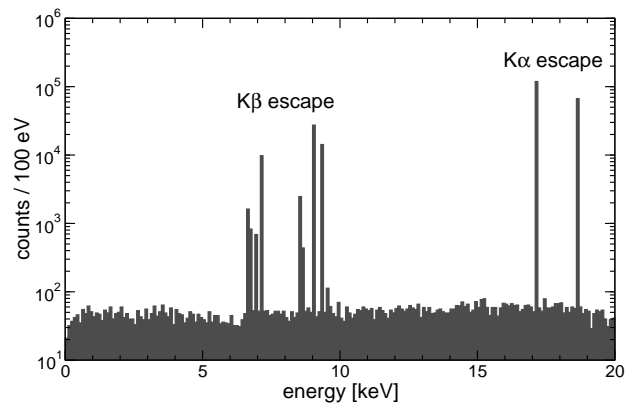}}
\caption{\label{fig:ti44simul} Simulation of the $^{44}$Ti spectrum showing the escape lines.}
\end{center}
\end{figure}
The spectrum in Figure\,\ref{fig:ti44simul} is a Montecarlo simulation of the exposure to a  \Ti\ source of a \agre\ detector
with size similar to the ones considered here. It is possible to recognize the
peaks due to the escape of \kaone, \katwo\ and \kb\ Re X-rays mentioned above:
the escape peaks between 5 and 10\,keV reflect the complex structure of the \kb\ X-ray line. The flat continuum is caused by Compton scattering of the \Ti\ $\gamma$-rays.

\begin{figure}
\resizebox{.45\textwidth}{!}{ 
\includegraphics*[]{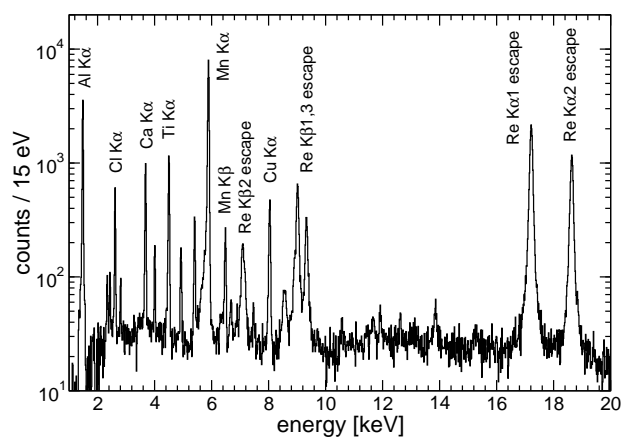}}
\caption{\label{fig:measured15} Summed spectrum of channels 5 and 6 from the RUN15 $^{44}$Ti measurement with the Re escape peaks. The  beta spectrum has been subtracted -- about $1.5\times10^5$\,counts between 1.0 and 2.465\,keV.}
\end{figure}

In RUN15 the $^{44}$Ti source could not be shielded and the detectors were always
exposed to its radiation. This caused a high level of background during the measurements with the fluorescence source (see Figure~\ref{fig:measured15}).  
Moreover, because of the intensity of the  $^{44}$Ti source
the analysis of RUN15 data is severely impaired by pile-up. Due to the
asymmetrical placement of the $^{44}$Ti source, two of the eight detectors
(detector 5 and 6) show a lower pile-up rate with a smaller impact on their
performance. Therefore, for the analysis discussed in this paper only these two detectors were used. Figure~\ref{fig:measured15} shows the energy spectrum obtained from the whole RUN15 measurement for these two detectors. The total measurement time is again about 1000\,hours. The FWHM energy resolution is about 31\,eV for the Cl K$_\alpha$ line, which is only 7\% worse than in RUN14. 

As discussed in the following, to investigate the Al and Cl peaks we have used also data from RUN9.
The total calibration time in this case amounts only to about 220 hours. Along with a lower statistics, the RUN9 spectrum has a slightly worse energy resolution with respect to RUN14. On the other hand, RUN9 is free from the low energy X-ray peaks from Pb fluorescence and gentler offline cuts did not affect the shape of the Al peak.

In the following we discuss the fitting of the experimental calibration spectra as well as the spectrum of the \ti\ measurements. Then we present the results of the peak shape investigation by means of Montecarlo simulations.

\section{Fitting of the Experimental Spectra}
Tails in experimental spectra have also been observed in standard ionization semiconductor
detectors. In particular, in alpha silicon detectors a tail is attributed to
energy loss fluctuations in the collisions of the alpha particles with the silicon nuclei in the crystal \cite{bla98}.
The resulting asymmetric peaks $T(E,E_0)$ are generally described as the convolution of a
normalized Gaussian $G(E,E_0)$ with an exponential $R(E,E_0)$ \cite{bla98,bor87}:
\begin{eqnarray}
G(E,E_0) & = & \frac{1}{\sigma \sqrt{2 \pi}} \exp \left[ - \frac{1}{2} \left( \frac{E - E_{0}}{\sigma} \right)^{2} \right] \\
R(E,E_0) & = & \lambda \exp \left[ (E - E_{0})
  \lambda \right] \cdot h(E_0-E) \\
T(E,E_0) & = & G(E,E_0) \otimes R(E,E_0) \\
& = & \frac{\lambda}{2} \exp \left[ (E - E_{0}) \lambda + \left( \frac{\sigma
      \lambda}{\sqrt{2}} \right)^2 \right] \nonumber \\
 & & \cdot \mathrm{erfc} \left[ \frac{E - E_{0}}{\sigma \sqrt{2}} + \frac{\sigma \lambda}{\sqrt{2}} \right] \nonumber  
\end{eqnarray}
where $E_0$ is the peak position, $\sigma$ is the Gaussian standard deviation, $\lambda$ is the exponential tailing constant,  $h(E_0-E)$ is the Heavyside function, and erfc is the complementary error function. 

The above function $T(E,E_0)$ describes one Gaussian peak with an exponential tail. However, we found that this simple approach does not provide a satisfactory description of the lines in our spectra (Figure~\ref{fig:measured}). In our case, the best description is given by 
the sum of three independent components: one main symmetric Gaussian peak $G(E,E_0)$ and two asymmetric Gaussian peaks like the above $T(E,E_0)$.
The two asymmetric peaks are the convolution of the Gaussian $G(E,E_0)$ with two exponentials, $R_1(E,E_0)$ and $R_2(E,E_0)$, with tail parameters $\lambda_{1}$ and $\lambda_{2}$:
\begin{eqnarray} \label{fitfunc}
F(E,E_0) & = & G(E,E_0)\otimes[\delta(E) + R_1(E,E_0) + R_2(E,E_0)] \nonumber  \\ 
F(E,E_0) & = & \frac{A_{Gauss}}{\sigma \sqrt{2 \pi}} \exp \left[ - \frac{1}{2} \left( \frac{E - E_{0}}{\sigma} \right)^{2} \right] \\ 
 & + & A_{exp1} \frac{\lambda_{1}}{2}  \exp \left[ (E - E_{0}) \lambda _{1} +
   \left( \frac{\sigma \lambda _{1}}{\sqrt{2}} \right)^2 \right] \nonumber \\
 & & \cdot \mathrm{erfc} \left[ \frac{E - E_{0}}{\sigma \sqrt{2}} +
   \frac{\sigma \lambda_{1}}{\sqrt{2}} \right] \nonumber \\
 & + &  A_{exp2}  \frac{\lambda_{2}}{2} \exp \left[ (E - E_{0}) \lambda _{2} +
   \left( \frac{\sigma \lambda _{2}}{\sqrt{2}} \right)^2 \right] \nonumber \\
 & & \cdot \mathrm{erfc} \left[ \frac{E - E_{0}}{\sigma \sqrt{2}} + \frac{\sigma \lambda_{2}}{\sqrt{2}} \right] \nonumber 
\end{eqnarray}
$A_{Gauss}$ is the area of the Gaussian peak, $A_{exp1}$ and $A_{exp2}$ describe the amplitudes of the two exponential tails.
As will be argued in the following, one of the two exponential tails can be explained as a consequence of the shallow X-ray absorption depth. 

In spite of the relatively large width of the Gaussian peaks (ranging from
about 25 to about 45\,eV FWHM) a precise description of the measured spectra
calls for a detailed modeling of the X-ray emission. This is particularly
true for the Re \ka\ escape peaks whose natural line width $\Gamma$  -- about
47\,eV -- is comparable to the Gaussian line width $\sigma$ -- about 28\,eV. Therefore, both the
doublet structure of the K$\alpha$ lines and the natural widths $\Gamma$ of X-ray transitions have to be included. 

To take into account the natural line widths, for each peak the above response function $F(E,E_0)$ was numerically convoluted with an asymmetric Breit-Wigner formula \cite{don70}:
\begin{eqnarray}
BW(E, E_0, \Gamma, \delta_{AS}) & = &\frac{\delta_{AS}}{\pi(1+\delta_{AS})}
\frac{\Gamma _{L}}{(E - E_{0})^{2} + \Gamma _{L}^{2}/4} \nonumber \\ 
\mbox{ for } E \leq E_{0} \\
BW(E, E_0, \Gamma, \delta_{AS}) & = & \frac{1}{\pi(1+\delta_{AS})}\frac{\Gamma
  _{R}}{(E - E_{0})^{2} + \Gamma _{R}^{2}/4} \nonumber \\ 
\mbox{ for } E > E_{0} \\ 
\Gamma _{R} & = & \frac{2 \Gamma}{1 + \delta _{AS}}, \: \Gamma _{L} = 2 \Gamma - \Gamma _{R}
\end{eqnarray}
Hereby, $\Gamma$ is the natural line width and $\delta _{AS}=\Gamma_L/\Gamma_R$ the asymmetry index. The values used for the fits are listed in Table\,\ref{tab:parameters}. It is worth noting that it is not quite straightforward
to find reliable atomic parameters in the literature, especially for the low Z atoms like Al and Cl.
Moreover, line widths and asymmetry indexes depend on the chemical form of materials \cite{kaw86,kaw89}. 
Whenever possible we have used the most recent and appropriate experimentally determined parameters. 
Lacking a direct experimental determination, the natural widths of Cl and Re lines 
are the ones recommended in a recent compilation of experimental and theoretical values. 
When published values were missing -- as for Al, Cl and Re -- the asymmetry index has been arbitrarily set to 1.0.
Given all the above, some systematic uncertainty on the fit parameters has to be expected.

\begin{table}[b]
\begin{center}
\begin{tabular}{lccccc} \hline
line & $\Gamma$ [eV] & $\delta_{AS}$ & $\frac{I_{K \alpha 2}}{I_{K \alpha 1}}$ & $\Delta E$ [eV] & ref.\\ 
\hline 
Mn \ka1 & 2.47 & 1.57 & \multirow{2}[0]*{0.510} & \multirow{2}[0]*{11.2} & \multirow{2}[0]*{\cite{hol97}}\\
Mn \ka2 & 2.92 & 1.26 &  &  & \\
Ti \ka1 & 1.87 & 1.19 & \multirow{2}[0]*{0.508} & \multirow{2}[0]*{5.96} & \multirow{2}[0]*{\cite{par36}}\\
Ti \ka2 & 2.34 & 0.98 &  &  & \\
Ca \ka1 & 0.98 & 1.15 & \multirow{2}[0]*{0.506} & \multirow{2}[0]*{3.56} & \multirow{2}[0]*{\cite{kaw89}\cite{par36}$^{*}$}\\
Ca \ka2 & 0.98 & 1.13 &  &  & \\
Cl \ka1,2 & 0.72 & 1.0 & 0.505 & 1.61 & \cite{kra79}\\
Al \ka1,2 & 0.85 & 1.0 & 0.503 & 0.43 & \cite{citr74}\\ 
Re \ka1 escape & 47.20 & 1.0 & \multirow{2}[0]*{0.580} & \multirow{2}[0]*{1422.4} & \multirow{2}[0]*{\cite{kra79}}\\
Re \ka2 escape & 47.60 & 1.0 &  &   & \\
\hline 
\end{tabular}
$^{*}\Gamma$ taken from \cite{kaw89}, other parameters from \cite{par36}
\caption{\label{tab:parameters} The parameters $\Gamma$ (natural linewidth),
  $\delta_{AS}$ (asymmetry index), $\frac{I_{K \alpha 2}}{I_{K \alpha 1}}$
  (ratio of intensities between \kaone\ and \katwo) and $\Delta E$ (energy
  difference between \kaone\ and \katwo) used in the fits.}
\end{center}
\end{table}

The Re \ka1 and \ka2 escape lines and the \kb\ lines -- whose structure is too fine and complex to be appreciable -- can
be satisfactorily fitted with the single peak $H_s(E,E_0)$ fit function:
\begin{equation} \label{fitfunc2}
H_s(E,E_0) = F(E,E_0) \otimes BW(E, E_0, \Gamma, \delta_{AS}) + background
\end{equation}
For the other \ka\ lines, the \ka1-\ka2 structure -- although not fully resolved by our detectors -- must be
accounted for by the doublet fit function $H_d(E,E_0)$:
\begin{eqnarray} \label{fitfunc3}
H_d(E,E_0) & = & F(E,E_0) \otimes[BW(E, E_{K \alpha 1}, \Gamma_{K \alpha 1}, \delta_{AS,K \alpha 1}) \nonumber  \\ \nonumber 
& + & \frac{I_{K \alpha 2}}{I_{K \alpha 1}} \cdot BW(E, E_{K \alpha 1} -
\Delta E, \Gamma_{K \alpha 2}, \delta_{AS,K \alpha 2})] \nonumber \\ 
& + & background
\end{eqnarray}
The energy difference $\Delta E$ and relative intensities of the \ka\ lines $I_{K \alpha 1}$ and $I_{K \alpha 2}$ are listed in Table\,\ref{tab:parameters} as well.
Since tails from peaks at higher energies add up to an approximately constant background at lower energies, a constant background parameter was added in both $H_s(E,E_0)$ and $H_d(E,E_0)$. 
The single and doublet peaks of the summed spectra in Figure\,\ref{fig:measured} were fitted individually using $H_s(E,E_0)$ and $H_d(E,E_0)$, respectively.
The free fit parameters are the peak position $E_0$ (or $E_{K \alpha 1}$ for doublets), the three amplitudes $A_{Gauss}$, $A_{exp1}$ and $A_{exp2}$, the Gaussian standard deviation $\sigma$, the exponential tail parameters $\lambda_1$ and $\lambda_2$ and the constant background level. 
Because of the low statistics and the large number of fit parameters, the \kb\
peak fit results have not been considered. 

\begin{figure*}
\begin{minipage}{\textwidth}
 \resizebox{.48\textwidth}{!}{\includegraphics*[]{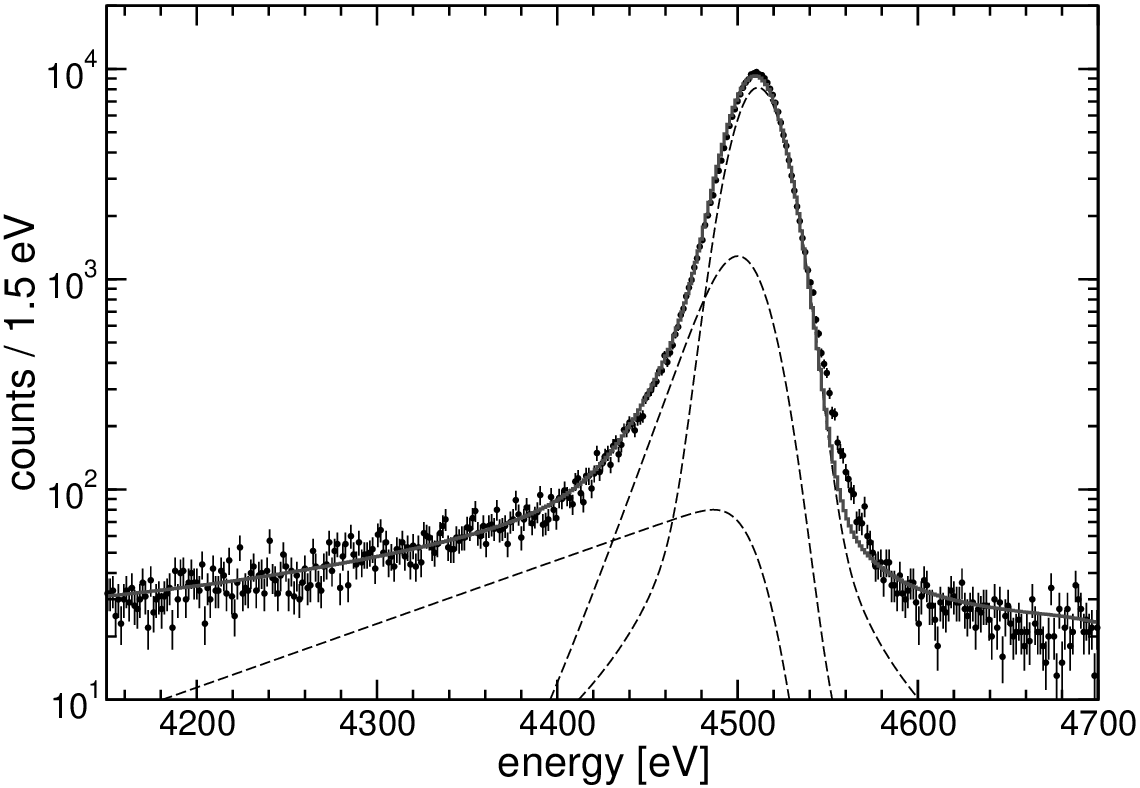}}
\resizebox{.48\textwidth}{!}{\includegraphics*[]{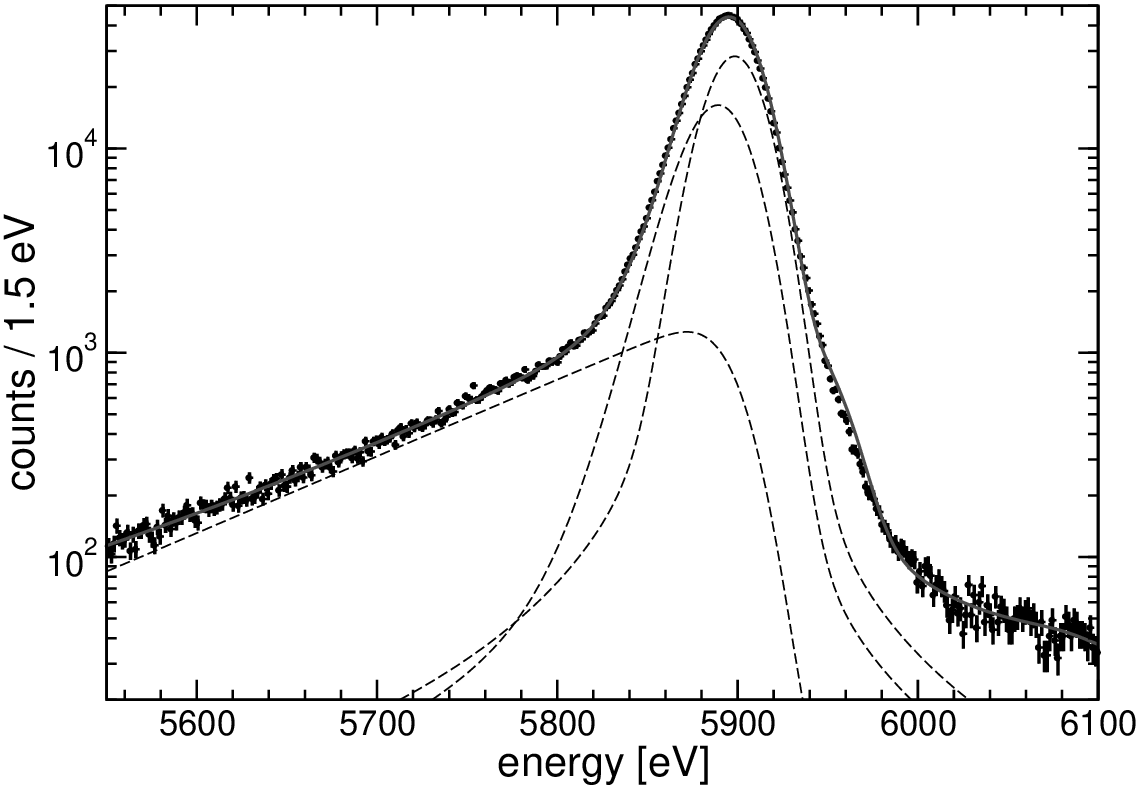}}
\caption{\label{fig:fitsum} Two examples for fits of the summed spectrum are displayed: Ti \ka\ (left side) and Mn \ka\ (right side). The dashed lines are the three components in equation (\ref{fitfunc}).}
\end{minipage}
\end{figure*}

Figure~\ref{fig:fitsum} shows as examples the fits of the Ti \ka\ and the Mn
\ka\ peaks for the RUN14 summed spectrum. The bump on the right side of the Mn peak is caused by the \kb\ peak of Cr at 5.950\,keV, also included in the fit.

\begin{figure*}[p]
\begin{minipage}{\textwidth}
\resizebox{.45\textwidth}{!}{ \includegraphics*[]{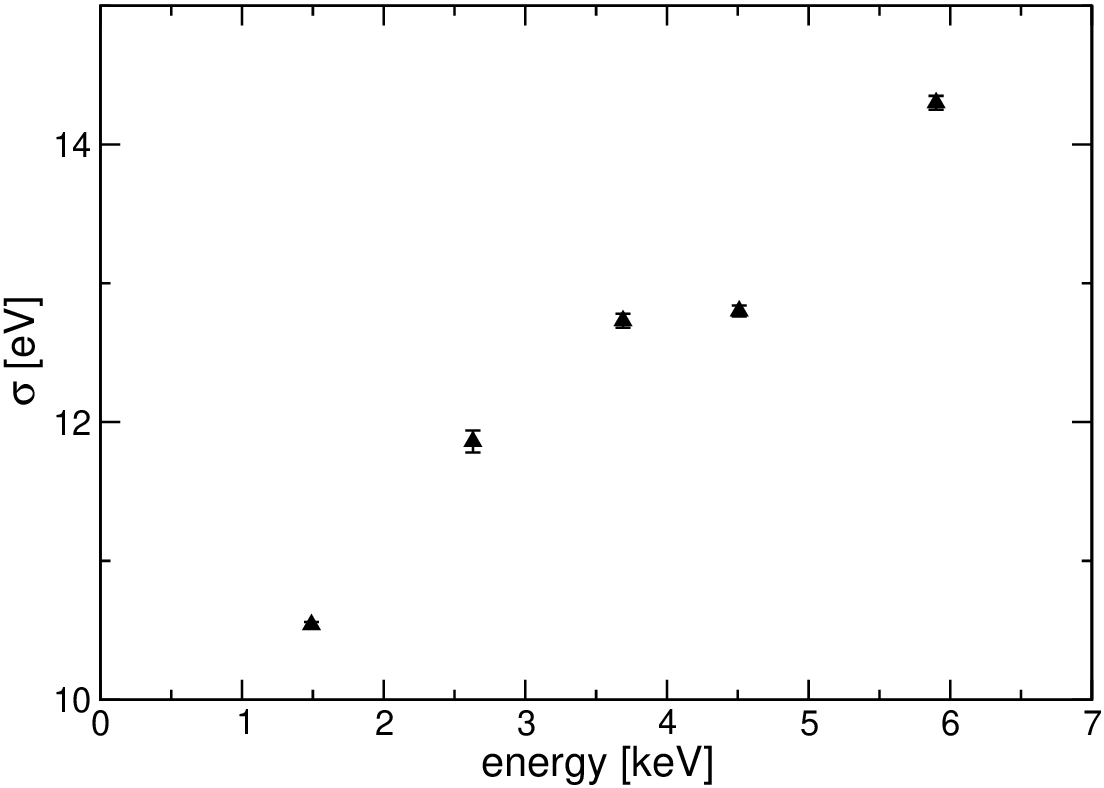}}
\hfill
\resizebox{.45\textwidth}{!}{ \includegraphics*[]{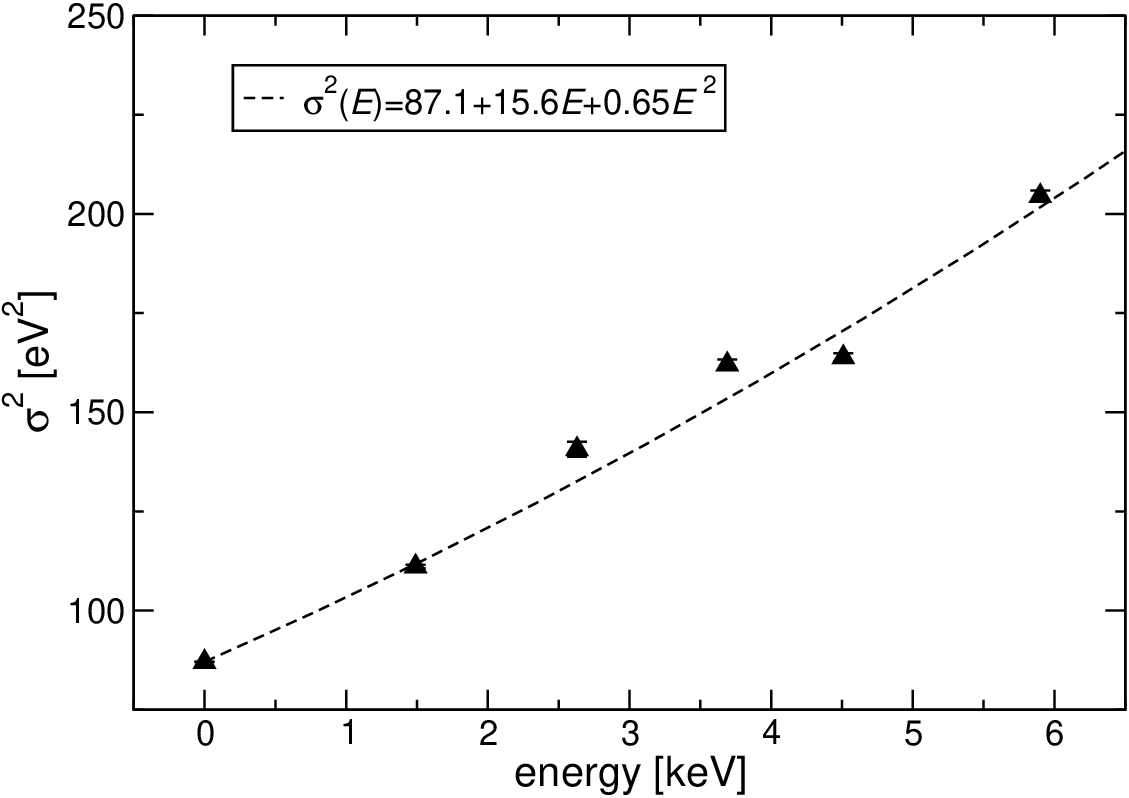}}
\caption{\label{fig:sigma}\label{fig:sigma2} The fit parameter $\sigma$ is displayed on the left side.
On the right side $\sigma^ {2}$ is displayed in
  dependence on the X-ray energy.}
\end{minipage}
\begin{minipage}{\textwidth}
\resizebox{.45\textwidth}{!}{ \includegraphics*[]{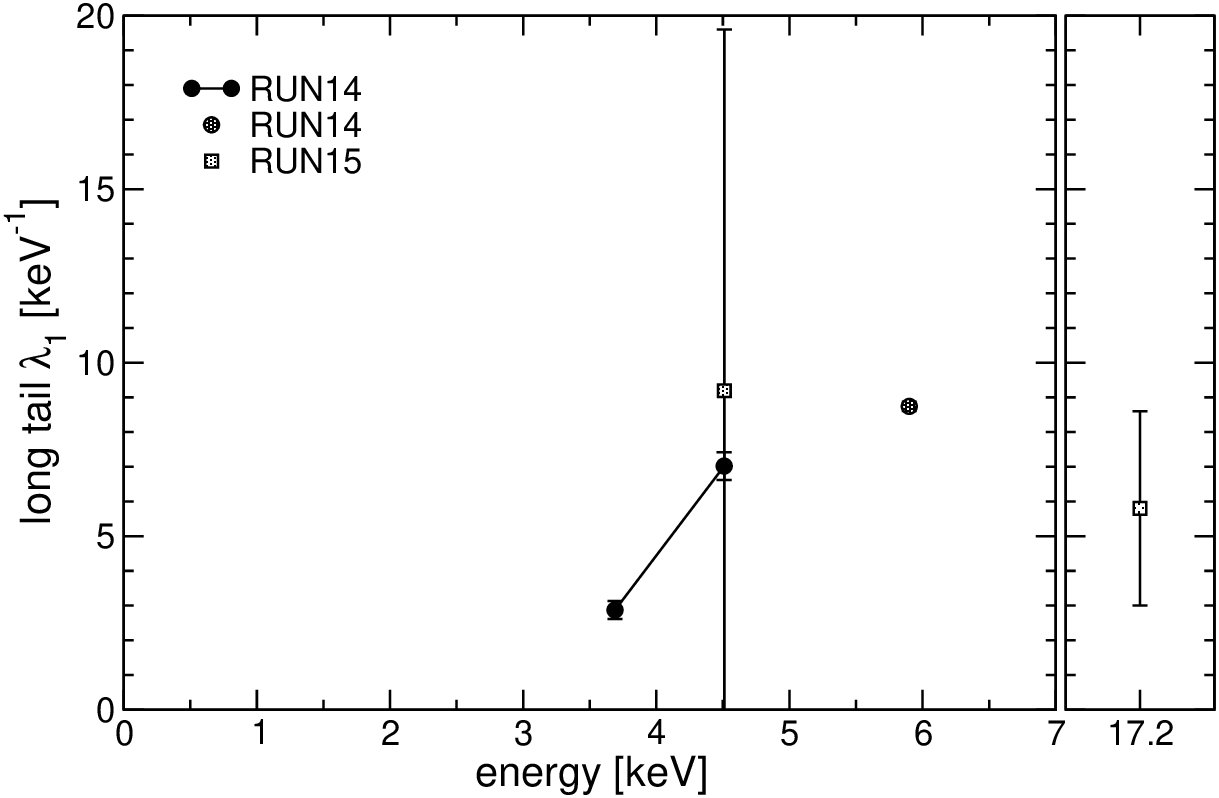}}
\hfill
\resizebox{.45\textwidth}{!}{ \includegraphics*[]{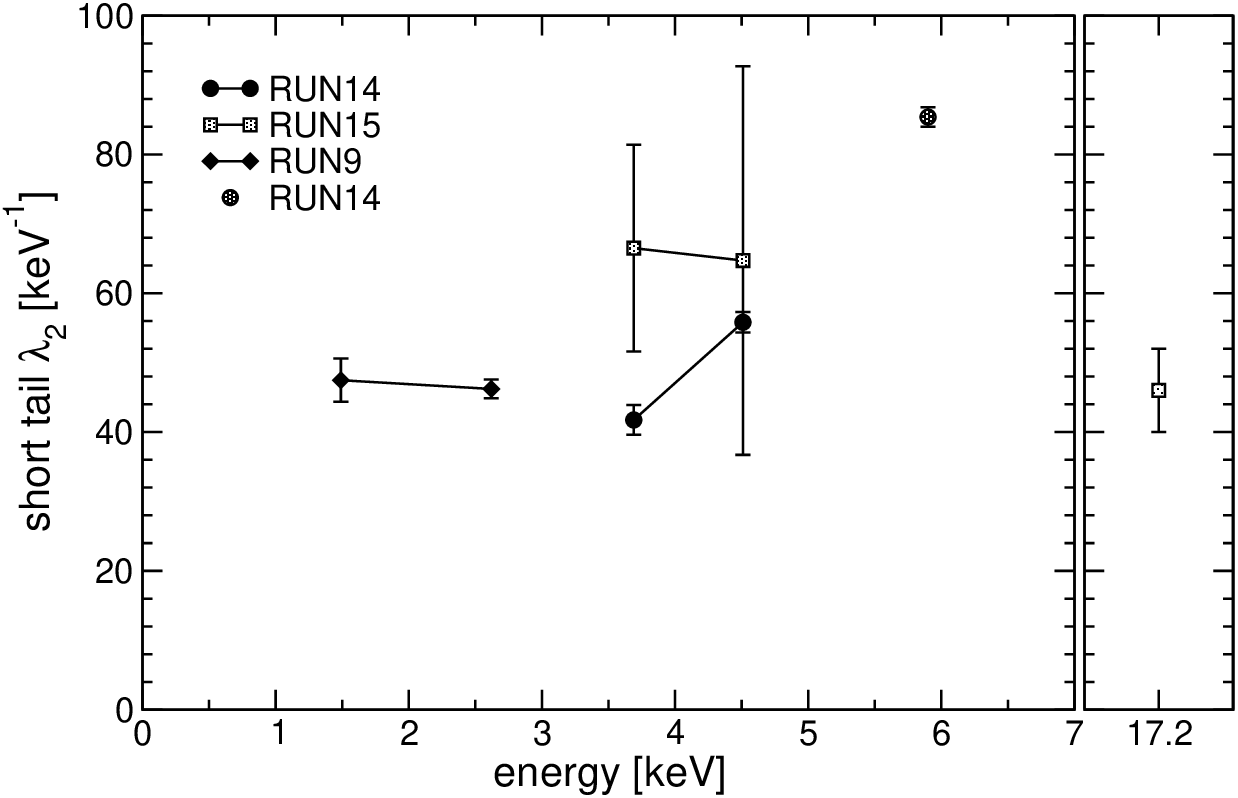}}
\caption{\label{fig:lambda} The fit parameters $\lambda_1$ and $\lambda_2$ are
  displayed in dependence on the X-ray energy. Filled circles and open
  squares are the RUN14 and RUN15 data, respectively.}
\end{minipage}
\begin{minipage}{\textwidth}
\resizebox{.45\textwidth}{!}{ \includegraphics*[]{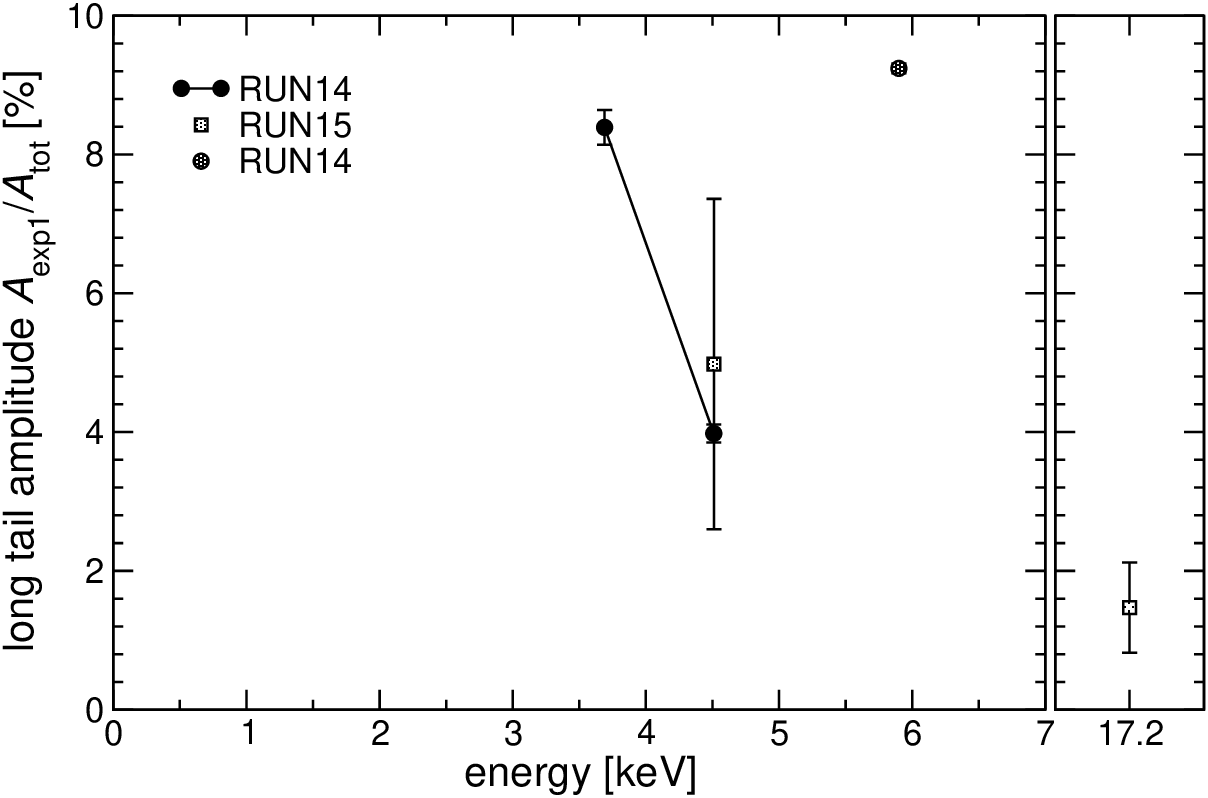}}
\hfill
\resizebox{.45\textwidth}{!}{ \includegraphics*[]{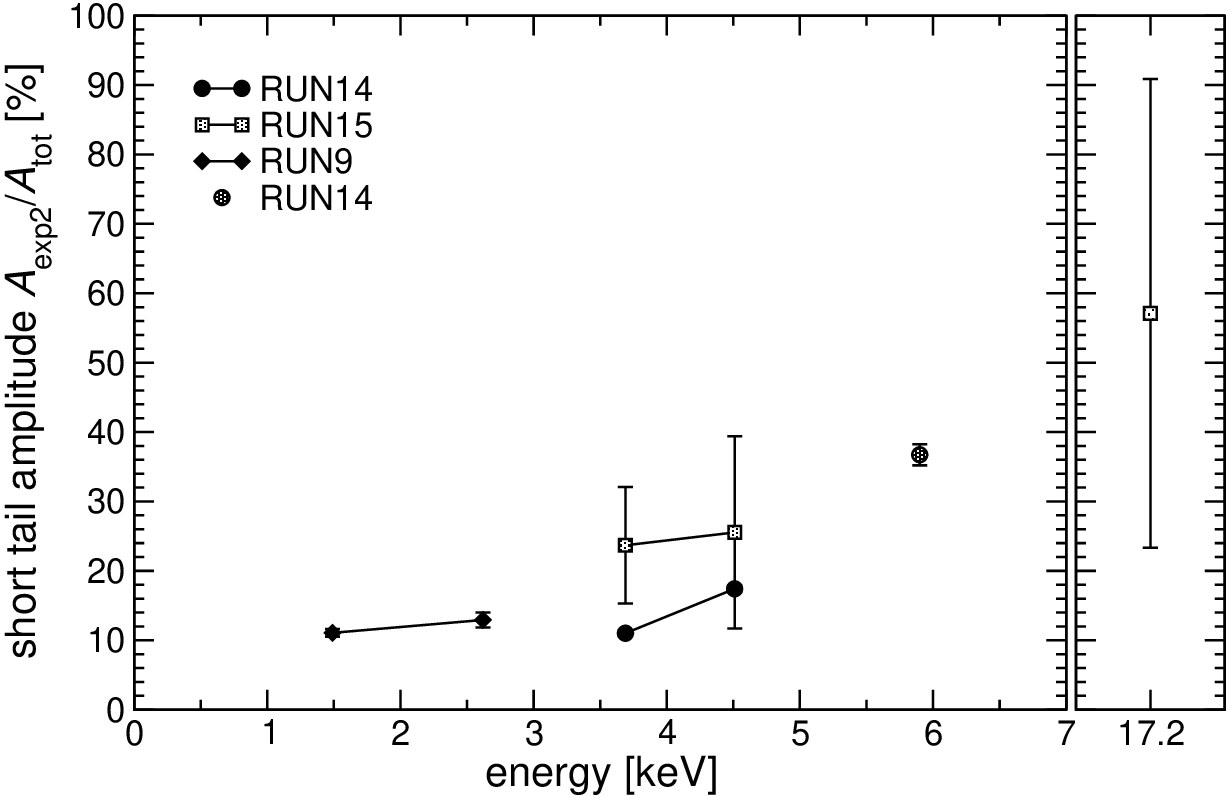}}
\caption{\label{fig:aexp} The fit parameters A$_{exp1}$ and A$_{exp2}$ are
  displayed in dependence on the X-ray energy.  Filled circles and open
  squares are the RUN14 and RUN15 data, respectively.}
\end{minipage}
\end{figure*}

Figure\,\ref{fig:sigma},\,\ref{fig:lambda} and\,\ref{fig:aexp} show the results for the parameters $\sigma$, $\lambda_1$, $\lambda_2$, $A_{\exp1}$ and $A_{\exp2}$ as found by
fitting the \ka\ peaks  in the calibration spectra. In these plots the tail amplitudes $A_{exp1}$ and $A_{exp2}$ are normalized to the total peak area $A_{tot}=A_{Gauss}+A_{exp1}+A_{exp2}$.

As far as the tail parameters are concerned, Figure\,\ref{fig:lambda} and \ref{fig:aexp} show the fit results only for a subset of
the peaks we have analyzed. These results were selected considering the stability of the fit procedure with respect to the initial parameter values and to the energy interval as well as the convergence to meaningful values of all free parameters.

According to this considerations, a clean analysis of RUN14 data is possible only for the Ca, Ti and Mn peaks.
As for the Cl peak, in the RUN14 spectrum the two
M Pb peaks are too close for a correct estimation of the tail parameters (see
spectrum in Figure\,\ref{fig:measured}), so the RUN9 spectrum was used
instead. However, the low statistics and high background in this measurement
prevent from properly estimating the long tail parameters, which are
accordingly omitted in Figure\,\ref{fig:lambda} and \ref{fig:aexp}. For the Al
peak in RUN14, the determination of the long tail parameters is not possible because the peak shape has been altered by the
software cuts applied to the data in order to separate spurious events from
true beta decays (see \cite{MIBETA} for details about the data
analysis). Although in the analysis of RUN9 gentler offline cuts did not
affect the shape of the Al peak, the long tail parameters are negatively
influenced by the presence of the beta spectrum. Therefore, they are also omitted in Figure\,\ref{fig:lambda} and \ref{fig:aexp}.

The results of the fits for the $\sigma$ parameter are displayed in
Figure\,\ref{fig:sigma}. The $\sigma$ parameter is the only one showing an
evident trend with energy. Its energy dependency may be described by a sum of
three squared contributions $\sigma^2(E)=a+bE+cE^2$, where the constant term $a$ is the squared baseline width, i.e. the
electronic noise contribution. We attribute the $b$ term to statistical
fluctuations ($\propto \sqrt{E}$) in the thermalization of the AgReO$_4$ and the  $c$ term to uncorrected gain instabilities ($\propto E$).

The fit results for the other parameters are summarized in
Figure\,\ref{fig:lambda} and \ref{fig:aexp}. Concerning the Mn peak, a Compton
back-scattering tail, as discussed in the previous section, accounts almost
completely for the observed long tail. All other parameters are compatible with the hypothesis of no energy
dependency. The safest conclusion is that, if any dependency exists at all, it
is hidden by statistical and systematic errors or by the cross-correlations between the various parameters.

\begin{figure}
\begin{center}
\resizebox{.45\textwidth}{!}{\includegraphics*[]{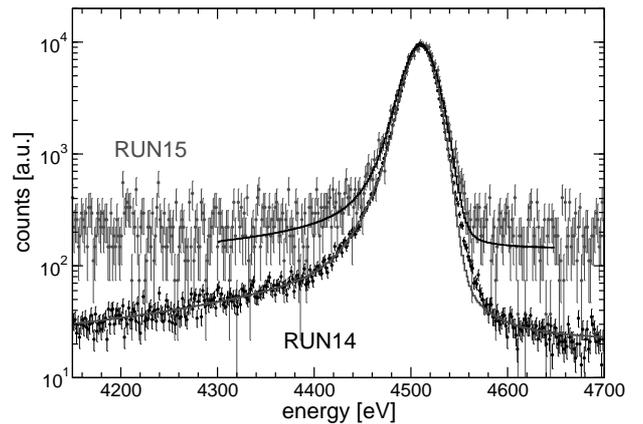}}
\caption{\label{fig:fitesc} Comparison of the normalized Ti \ka\ peaks in RUN14 (black) and RUN15 (grey).}
\end{center}
\end{figure}

Because of the lower signal-to-background ratio in the RUN15 sum calibration spectrum, a satisfactory fit was 
possible only for the Ti \ka\ peak. As can be seen in Figure\,\ref{fig:lambda} and\,\ref{fig:aexp},
both the long and short tail parameters are found to be compatible with the
ones of RUN14. This is also illustrated in Figure~\ref{fig:fitesc} which shows
the Ti \ka\ peaks in RUN14 and RUN15, normalized to overlay.

\begin{figure*}
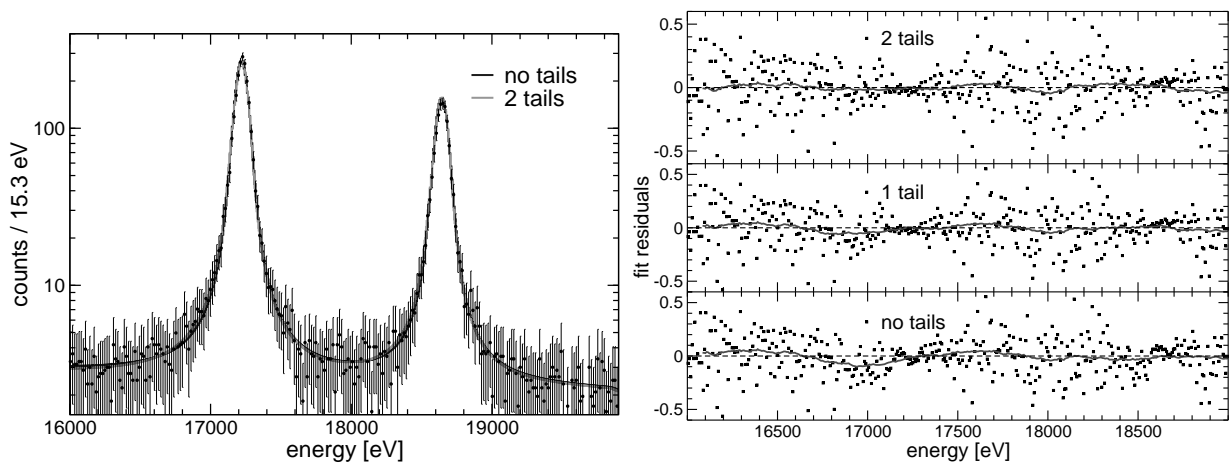

\resizebox{.45\textwidth}{!}{ \includegraphics*[]{fit-esc12_2.eps}}
\resizebox{.44\textwidth}{!}{ \includegraphics*[]{fitdiff-escKA12.eps}}
\caption{\label{fig:fitesc3} The simultaneous fit of the \kaone\
  and \katwo\ escape peaks with no tails and two tails is shown on the left
  side. On the right side, the corresponding fit residuals are displayed.}
 
\end{figure*}

For what concerns the escape peaks in RUN15, the analysis has been restricted to the Re \ka\ escape peaks since the \kb\ escape peaks show a too complex structure (see Figures\,\ref{fig:ti44simul} and \ref{fig:measured15}).
The Re \ka\ escape peaks have been fitted fixing their natural
widths $\Gamma$ and asymmetry indexes $\delta_{AS}$, according to the data in  Table\,\ref{tab:parameters}.  
Fits have been performed with the two tails  as in (\ref{fitfunc2}) as well as 
with one or no tail, both on the \ka1 alone and on the two \ka\ peaks together.
The results are listed in Table\,\ref{tab:escfitresults} and shown in Figure\,\ref{fig:fitesc3} for the two \ka\ peaks.
\begin{table}
\begin{tabular}{c|ccccc} 
\hline
\multicolumn{6}{c}{\kaone\, and \katwo\, escape peaks}\\
\hline
Peak& $\frac{A_{exp1}}{A_{tot}}$ & $\lambda_1$ & $\frac{A_{exp2}}{A_{tot}}$&$\lambda_2$ & $\chi^2$ \\
model & [\%] & [keV$^{-1}$] & [\%] & [keV$^{-1}$] & \\
\hline
no tail & -- & -- & -- & -- & 1.408\\
1 tail & -- & -- & $31\pm14$ & $35\pm6$  & 1.224\\
2 tails &  $1.5\pm0.7$ & $5.8\pm2.8$ & $57\pm34$ &$46\pm6$ & 1.210\\
\hline
\end{tabular}
\caption{\label{tab:escfitresults} Fit results for the two
   \kaone \ and \katwo \ escape peaks. }
\end{table}

Although the fits with the three models are almost undistinguishable by eye,
the $\chi^2$ suggests that the one without tails is the least satisfactory. 
The two tail and one tail model fits have very similar $\chi^2$, although the best fit is obtained with the two tail model.

The results for the  \ka1 escape peak fit are compared with the ones for the low energy peaks in Figure\,\ref{fig:lambda} and \ref{fig:aexp}. 
The comparison confirms that the $\lambda$ parameters do not depend strongly
on the energy. Considering instead the tail amplitudes, it may be
concluded  that, for the uniformly distributed escape events, whereas the long tail is clearly reduced -- as expected for a surface effect --,
the short tail is considerably higher than for the low energy peaks. However, it is worth noting that the short tail amplitude determination 
is hindered by the large escape peak width together with the limited statistics, as shown by the large error in Table\,\ref{tab:escfitresults}.

\clearpage

\section{Montecarlo Simulations}

To understand the origin of the observed tails another Montecarlo simulation based on the Geant4 toolkit
was developed. Hereby each X-ray line was simulated
separately using the energies and line widths listed in Tables
\ref{tab:energies1} and \ref{tab:parameters}, respectively. 
An additional exponential tail  was added  to the left of the Mn lines according to the results shown in Figure\,\ref{fig:mc_nuovo}.
For the simulation of the Re \ka\  escape lines, the initial \ti\ $\gamma$-line at $E = 78.337$
keV was used; the escape lines are then automatically generated by the Montecarlo
simulation. To account for the energy resolution of the detector and compare with experimental data, the resulting energy spectra
were convoluted with a Gaussian, whose $\sigma$ was taken from the
fitting of the respective line. As the \sip\
crystals do not have a regular geometrical shape \cite{MIBETA,nuc02}, they were
approximated by a cylinder with a diameter of $400 \mu$m and a thickness of $200 \mu$m. 

In addition to the summed spectrum, the sum of the two channels 5 and 6 was
investigated separately, because only these two channels were used for
the escape peak analysis. 

The Montecarlo simulation describes creation, transport and interaction of the initial photo-electron created by the X-ray photon, of the secondary electrons as well as of the secondary X-rays. 
However, it does not describe the energy transfer from the electron(s) to the phonon system which ultimately produces the signal observed with  thermal detectors. 
Incomplete energy detection that creates a low energy tail can be caused by either incomplete energy transfer from the electrons to the phonon system, or by a reduced temperature signal due to phonon energy losses or energy deposition in thermally weakly-coupled volumes. 

As the Montecarlo simulation does not account for these different processes,
the energy $E_{dep}$ deposited by the electron(s) in each Montecarlo step was
multiplied by a weight function $w(\vec{x}) \leq 1.0$, where $ \vec{x}$ is the position at
which the energy deposition happens. 
\begin{figure*}
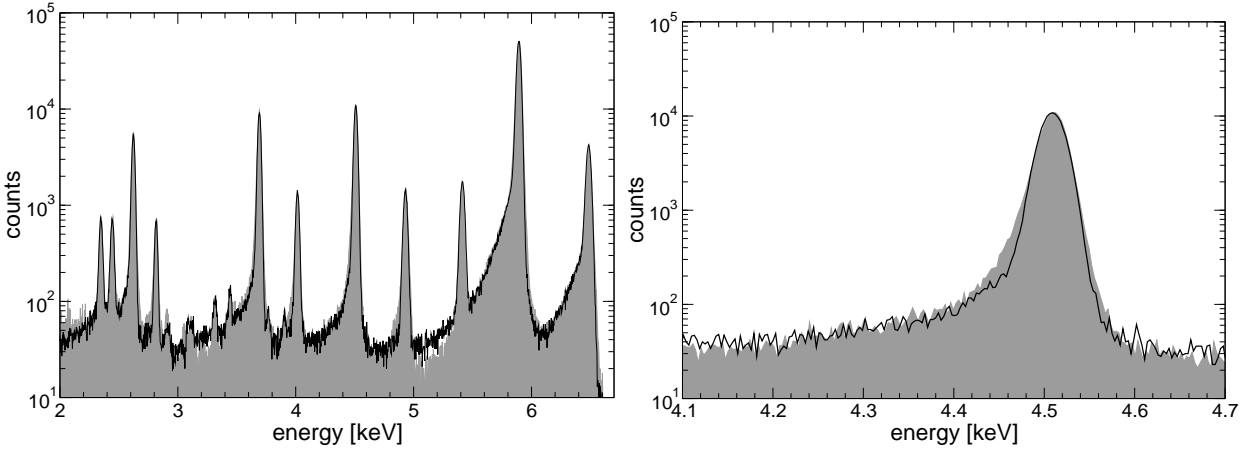
 
\resizebox{.45\textwidth}{!}{ \includegraphics*[]{sum_longshoulder.eps}}
\resizebox{.45\textwidth}{!}{ \includegraphics*[]{sum_longshoulderTi.eps}}
\caption{\label{fig:sim1} Comparison of the measured (grey) and simulated
  (black solid line) spectrum for the sum of eight detectors from the MIBETA measurement:
  full spectrum (left side) and Ti \ka\ calibration peak as a specific example
  (right side). Whereas the
  long shoulders are simulated sufficiently well, the short shoulders are not accounted for.}
\end{figure*}

As the fitting function could be divided into a long and a short shoulder, we have used a function $w(\vec{x})$ according to a model with two loss mechanisms.
The first one assumes that the \agre\ crystal has an insensitive surface layer of thickness $d_{0}$. The electron energy deposited in this layer does not contribute entirely to the thermal signal because of an incomplete conversion into phonons. The transition from insensitive to sensitive volume is smooth and described by a diffusion profile $f(d)$ with a diffusion length $DL_{d}$:
\footnote{The choice of this function clearly determines the exact tail shape. With the present choice the 
shape will be only approximately exponential, but the natural linewidth $\Gamma$, the noise $\sigma$ and the statistics prevent
from appreciating the difference.}
\begin{equation}\label{equ:diff1}
f(d) = \frac{1}{2} \cdot \left[ 1 + \mbox{tanh} \left( \frac{d - d_{0}}{DL_{d}} \right) \right]
\end{equation}
The result of this simulation is displayed in Figure\,\ref{fig:sim1} for the case of RUN14. 
The $d_{0}$ and $DL_{d}$ parameters were found by one-dimensional $\chi^2$ analysis to reproduce 
the experimental data in the 1.0 -- 7.0\,keV energy interval and are reported in Table\,\ref{tab:sim}.
It is shown clearly in  Figure\,\ref{fig:sim1}  that this simple model can describe the long shoulders sufficiently well,
although it fails to account for the short shoulders. 
\begin{table}
\begin{tabular}{lccc} \hline
run & RUN14 & RUN14 & RUN15 \\
spectrum & sum & channel 5+6 & channel 5+6 \\ \hline
$d_{0}$ [nm]& 20 & 52 & 52 \\
$DL_{d}$ [nm] & 64 & 52 & 52 \\
$r_{0}$ [$\mu$m] & 279 & 280 & 280 \\
$DL_{r}$ [$\mu$m] & 1 & 1 & 1 \\
L & 0.010 & 0.010 & 0.010 \\ \hline
\end{tabular}
\caption{\label{tab:sim} Optimal parameters for the simulations. The precision
  is $\Delta d_{0} = \pm 5$~nm, $\Delta DL_{d} = \pm 2$~nm, $\Delta r_{0} =
  \pm 1 \ \mu$m, $\Delta DL_{r} = \pm 100$~nm, $\Delta L = \pm 0.003$. The optimal values and the errors were
  determined by a one-dimensional $\chi^{2}$ analysis.}
\end{table}

\begin{figure*}
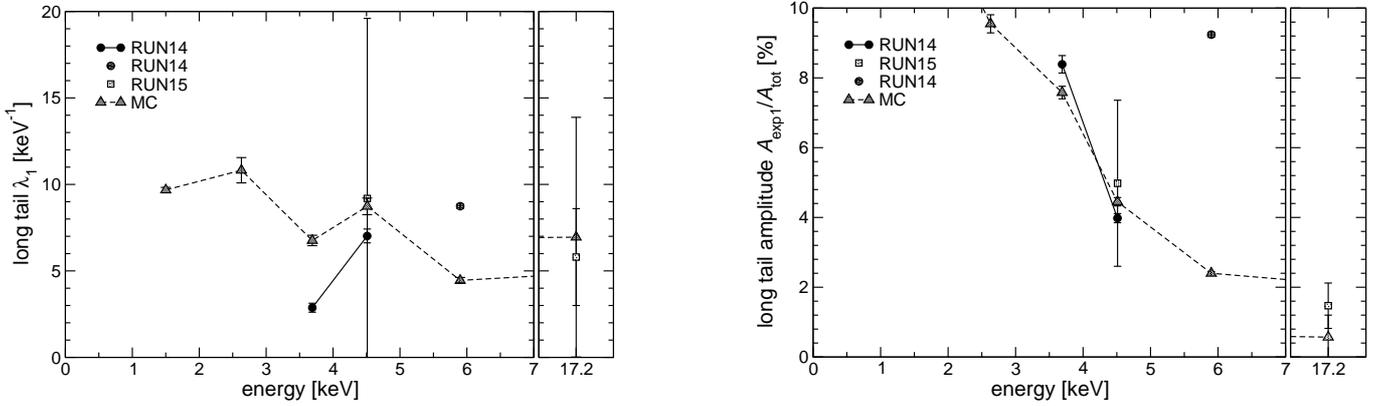
 
\begin{minipage}{\textwidth}
\resizebox{.45\textwidth}{!}{ \includegraphics*[]{lambda1_E-MC.eps}}
\hfill
\resizebox{.45\textwidth}{!}{ \includegraphics*[]{rap-A1_E-MC.eps}}
\caption{\label{fig:almc} The fit parameters A$_{exp1}$ and $\lambda_1$ are
  displayed in dependence on the energy of the X-rays.  Filled triangles are the MC data.}
\end{minipage}
\end{figure*}
The \ka\ peaks in the Montecarlo simulated spectrum (Figure\,\ref{fig:sim1}) were fitted with one single exponential tail and
the resulting long tail amplitudes $A_{exp1}$ and $\lambda_1$ parameters are plotted in Figure\,\ref{fig:almc} together with the results shown in
Figure\,\ref{fig:lambda} and \ref{fig:aexp}. The comparison confirms that this simple model indeed reproduces the features of the experimental data and,
as expected for a surface effect, the long tail disappears almost completely in the escape peaks.

As the short tail parameters $A_{exp2}$ and $\lambda_2$ show a completely different dependence on the energy,
another mechanism must be found to explain the short component of the experimental peaks.
Two types of effects can be the origin of the observed short tail. The first one consists of spatial effects, where the tail depends on the energy because the spatial distribution of the interaction does.
All effects which are not related to the interaction position belong to the second type. 

In fact there are good reasons to suspect the presence of volume effects that are caused by the dependence of the thermal signal on the
interaction position.
The calorimeter is a composite object with an absorber crystal and a
thermometer attached to it by means of a small glue spot.
If the phonons do not reach complete thermal
equilibrium in the whole absorber crystal faster than the detector thermal
time constant, part of them will be lost depending on the X-ray impact
location\footnote{Indeed the pulses observed with our \agre\ microcalorimeters
  cannot be understood only as thermal signals. In general they present two
  decay time constants. The longer one -- few tens of milliseconds -- could be interpreted as the detector thermal relaxation to the operating temperature. The shorter one -- few milliseconds -- is probably caused by the direct interaction of out-of-equilibrium phonons in the attached silicon thermometer.}. 
It is reasonable to assume that impact locations farther away
from the attachment point of the thermometer will create a reduced thermal
signal, thus creating a local dependence of the energy response. 

\begin{figure*} 
\epsfig{file=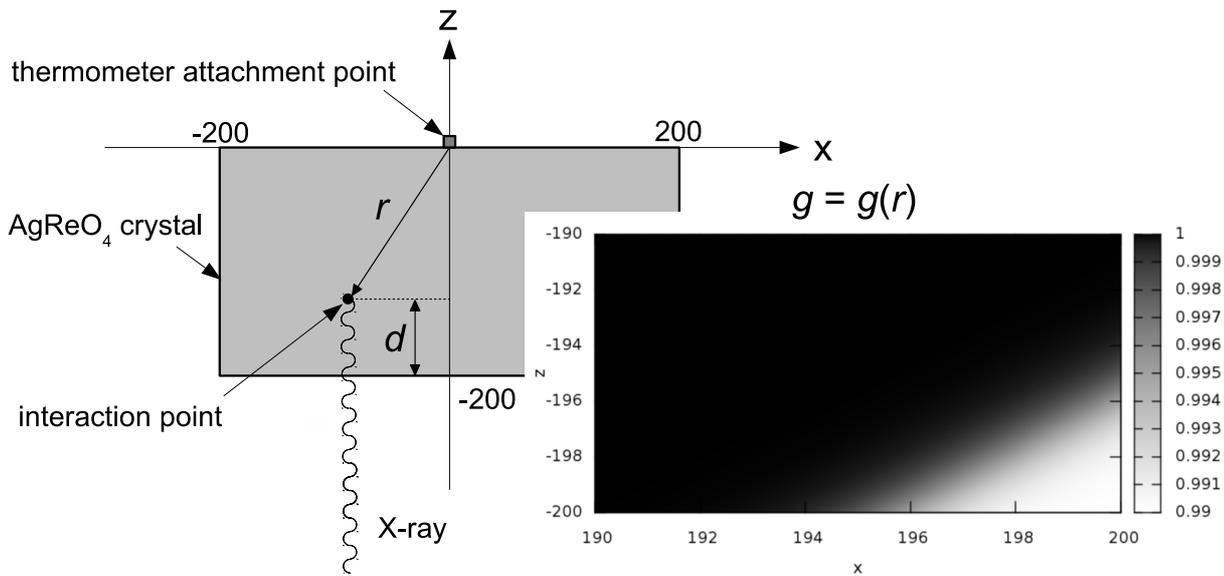,width=.9\textwidth}
\caption{\label{fig:det} Scheme of coordinates for Montecarlo simulations: $d$
  denotes the distance of the event from the detector surface, $r$ the distance
  from the glue point of the thermometer. See text for an explanation of $g(r)$ (equation \ref{equ:diff2}). Units are microns.}
\end{figure*}

Therefore, as an example for this loss mechanism, we report here on one model which has a dependence 
on the distance of the energy deposition from the attachment point of the thermometer. To account for a smooth
transition of sensitivity, we used a second diffusion model: 
\begin{equation}\label{equ:diff2}
g(r) = 1 - L \cdot \frac{1}{2} \cdot \left[ 1 + \mbox{tanh} \left( \frac{r - r_{0}}{DL_{r}} \right) \right]
\end{equation}
where $r$ stands for the distance of the impact location from the attachment point
(see Figure\,\ref{fig:det} for a schematic description),
$r_{0}$ is a ``thickness'' parameter describing the gradual decrease of
sensitivity for crystal regions farther away from the attachment point, and $DL_{r}$ describes the
transition width. $L$ sets the minimum $g$ value for the $r \gg r_0$ limit: 
 when $L = 1.0$, $g(r)$ approaches 0 for $r \gg r_0$.

The final weighting function $w(d,r)$ is then given by  
\begin{equation}\label{equ:diff3}
w(d,r) = f(d) \cdot g(r)
\end{equation}

\begin{figure*}[p]
\resizebox{.45\textwidth}{!}{ \includegraphics*[]{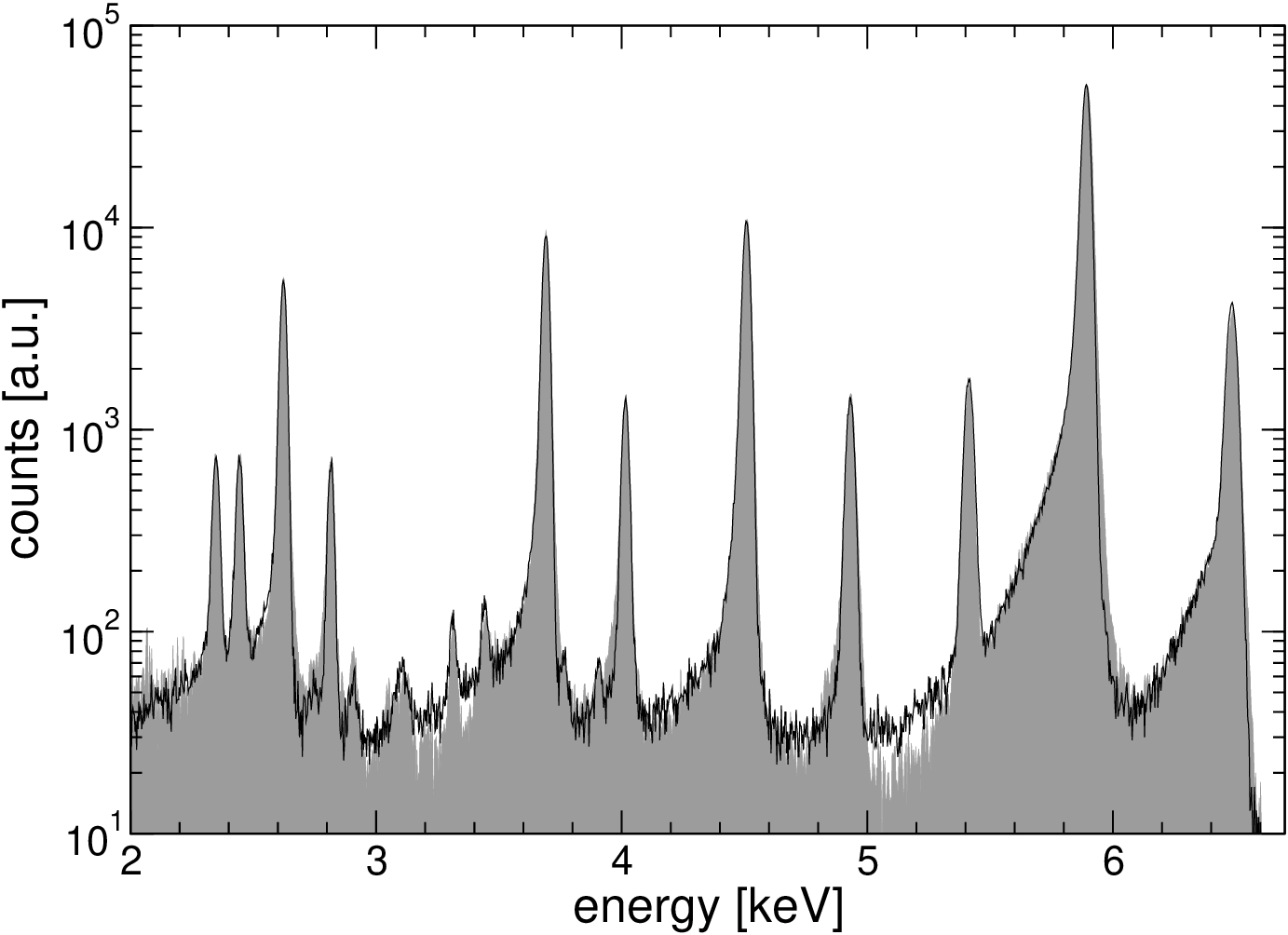}}
\resizebox{.45\textwidth}{!}{\includegraphics*[]{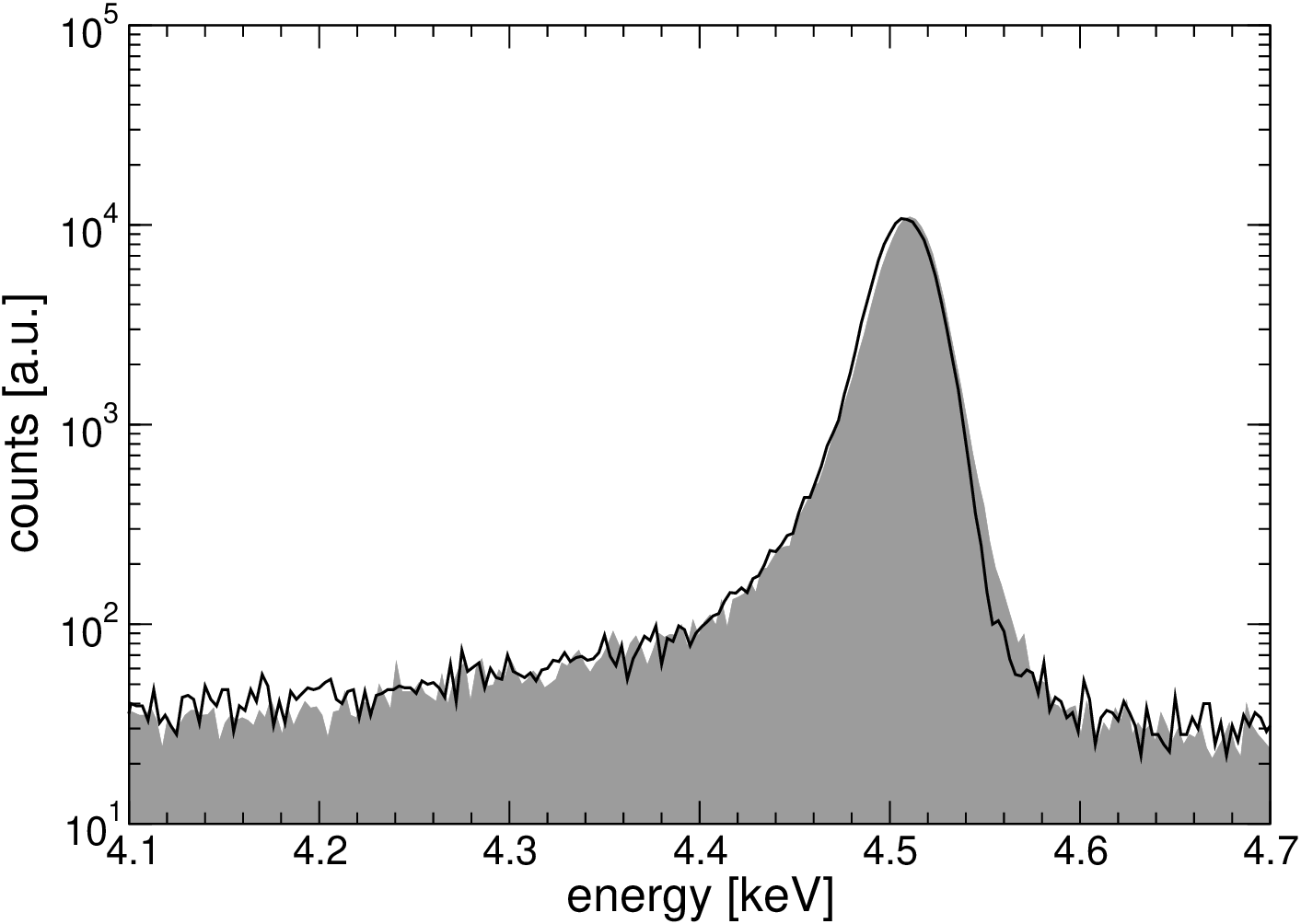}}
\caption{\label{fig:sim} Comparison of the measured (black) and simulated
  (grey) spectrum for the sum of eight detectors from the MIBETA measurement:
  the full spectrum on the left side, the Ti calibration peak as a specific
  example on the right side. For discussion see text.}
\end{figure*}

The parameters used in the simulations are summarized in Table \ref{tab:sim}: these parameters correspond to 
a weight function which is 1.0 in the whole crystal volume and decreases just on the lower rim of the cylindrical crystal (see Figure\,\ref{fig:det}).
Figure \ref{fig:sim} shows the result of the simulated spectrum compared with the measured spectrum from Figure~\ref{fig:measured}. 

From the values of the simulation parameters, it can also be deduced that the two
mechanisms have different origin. Whereas $d_{0}$ and $DL_{d}$ are rather
different for the sum of all and the sum of only two channels, this cannot be
said for the other parameters $r_{0}$, $DL_{r}$ and $L$. This strengthens our interpretation of the first
mechanism as a surface effect which is expected to be individual for each crystal. The similarity of the other parameters for the different channels points to a more systematic effect common to all crystals. 

\begin{figure*}[p] 
  \resizebox{.45\textwidth}{!}{ \includegraphics*[]{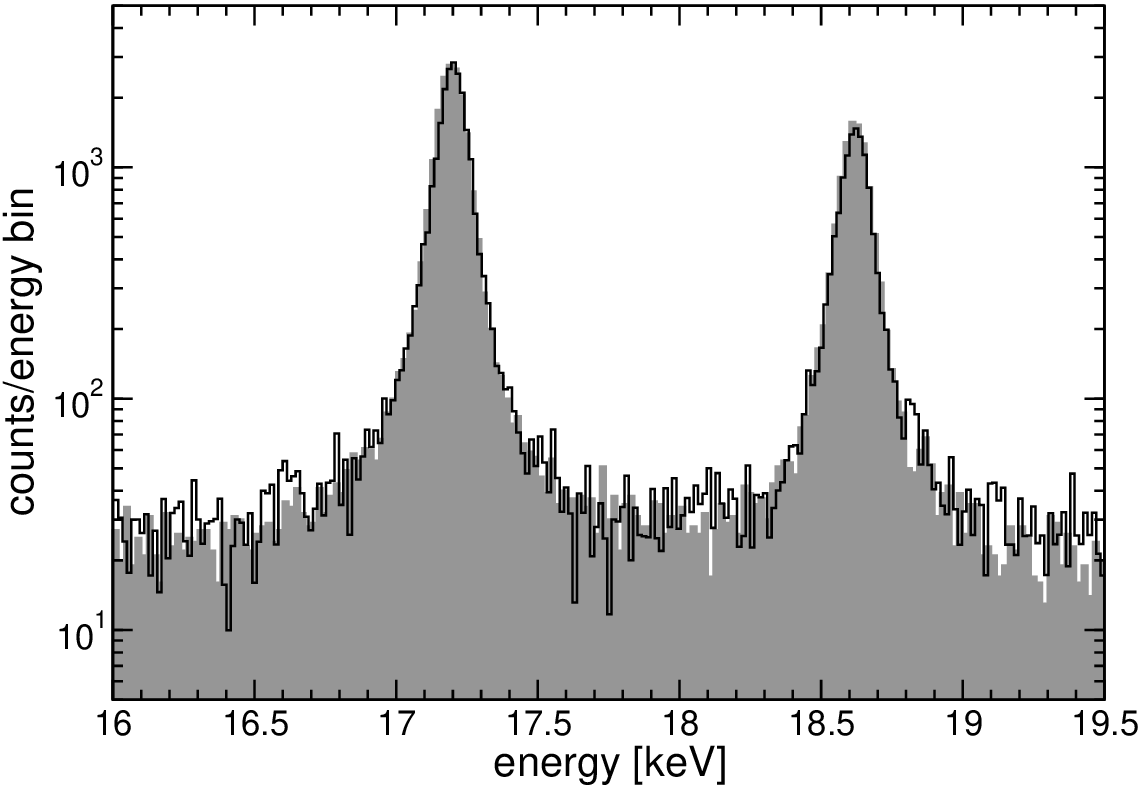}}
  \resizebox{.45\textwidth}{!}{\includegraphics*[]{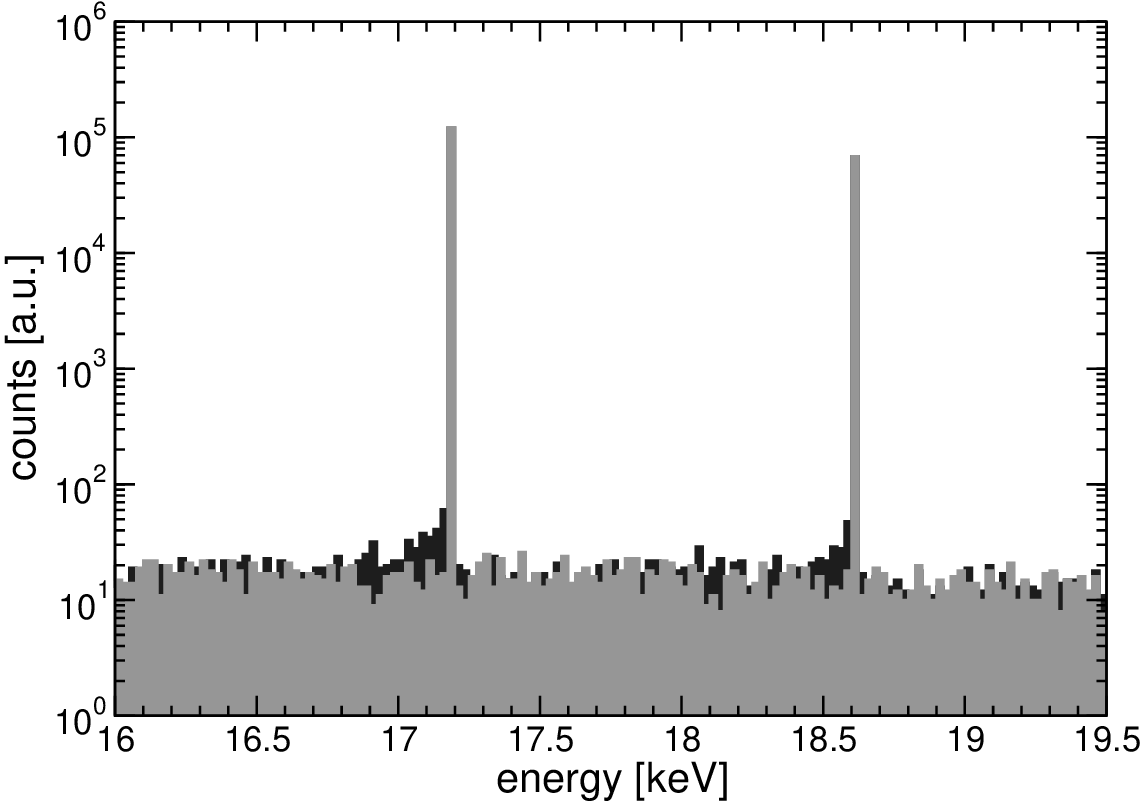}}
 \caption{\label{fig:sim2} The left panel shows the comparison of the
   measured (black) and simulated (grey) spectrum for the Re \kaone\ and \katwo\
   spectra. The right panel shows a spectrum without
   taking into account the Gaussian and natural line widths for a better
   estimation on a possible tail. The light and dark shaded histograms are with and
without the inclusion of the surface effect, respectively.}
 \end{figure*}

\begin{figure*}[p] 
  \resizebox{.75\textwidth}{!}{ \includegraphics*[]{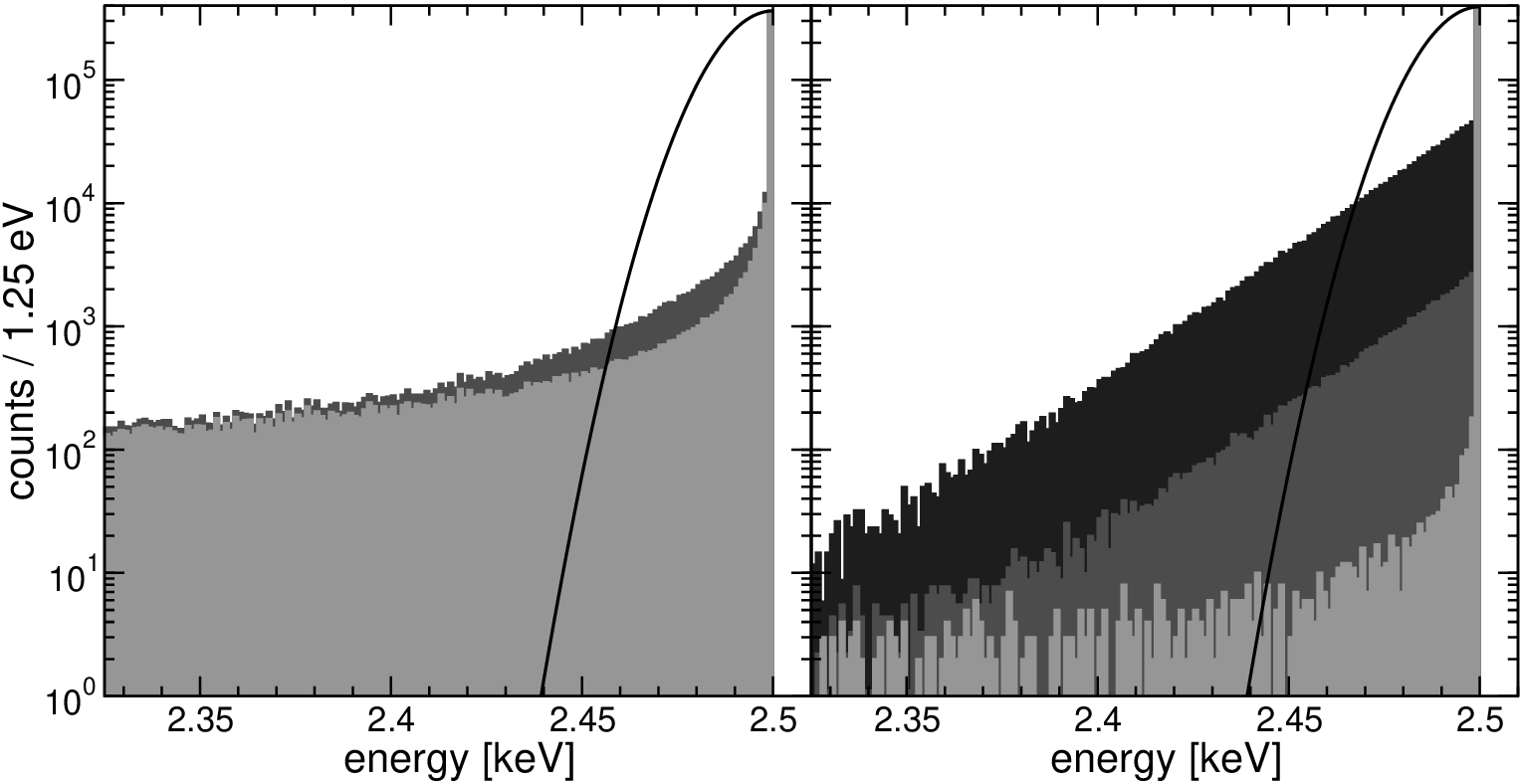}}
 \caption{\label{fig:beta} Simulation of the spectrum due to external X-rays 
    (right) and uniform internal $\beta$  (left) interactions. The energy is 2.5\,keV
in both cases. See text for color explanation.
}
 \end{figure*}

Figure \ref{fig:sim2} shows the measured as well as the simulated Re \ka\ escape peaks for the \ti\ measurement (RUN15).  
The Re \ka\ escape peaks were simulated with the same values for $r_{0}$, $DL_{r}$ and $L$ as for the calibration peaks, adding an increased low-energy background from the \ti\ source. The agreement between the experiment and Montecarlo as shown in 
the left panel of Figure\,\ref{fig:sim2} is fairly good. The right panel of
Figure\,\ref{fig:sim2} shows the escape peaks obtained with the Montecarlo
simulation when neither the transition natural width nor the detector
resolution are included: the plot shows that although no appreciable long term
tail is expected a very small short term component is indeed present in the simulation. However,
a comparison with the results in Table\,\ref{tab:escfitresults} and Figure
\ref{fig:aexp} shows that the Montecarlo fails to predict the correct amplitude
of this short tail by more than an order of magnitude. 

A model that depends on the distance of the interaction point from the calorimeter is
naturally more dependent on the geometrical shape of the crystal than a surface
model. However, other models for spatial dependence with different assumptions
for the geometry as well as for the weighting function $g(r)$ show similar
shortcomings.
   
\begin{figure} 
\resizebox{.45\textwidth}{!}{\includegraphics*[]{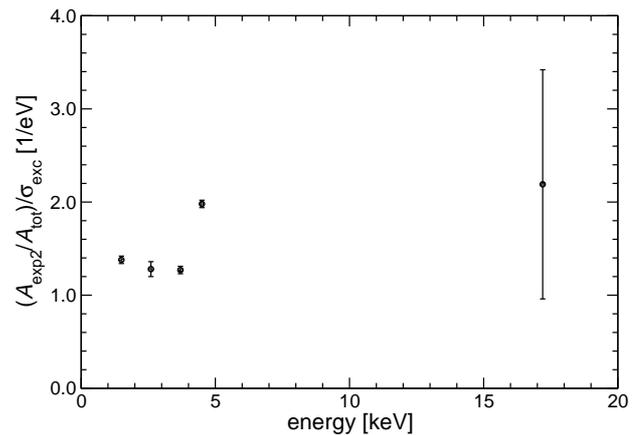}}
\caption{\label{fig:sl} Ratio of $A_{exp2}/A_{tot}$ to
  $\sigma^{2}_{exc} = \sigma ^{2} - \sigma _{0}^{2}$. }
\end{figure}

The second type of effects which can cause a tail may include, for example, 
complex energy losses due to trapping in metastable states, event pile-up as well as other analysis artefacts. 
In particular, for what concerns these effects, it is quite intriguing to notice that the short tail amplitude $A_{exp2}$ and the line width $\sigma$ vary with the energy in a similar way. This is illustrated in Figure\,\ref{fig:sl} where the ratio between the relative amplitude $A_{exp2}/A_{tot}$ and the peak broadening in excess of the baseline width ($\sigma_{exc}^2=\sigma^2-\sigma_0^2$) is plotted versus the energy. A common energy dependence might be inferred from this plot. This points to a possible common explanation for the two peak parameters.   

Indeed both the above types of effects could be present in our
detectors. However, with our current understanding it is not possible to build
models that satisfactory explain the short tail in our data. 
A better understanding of the mechanisms causing that short tail would require a higher statistics measurement with the $^{44}$Ti source
as well as calibrations with lower energy (i.e. $\leq 17$\,keV) $\gamma$ or X-ray sources to better investigate the tail energy -- and therefore spatial -- dependence.

The important aim of the present investigation was to determine to which
extent the tails observed in the calibration peaks affect the beta spectrum as
well. Unfortunately nothing conclusive was found for the short tail. 
Therefore in order to evaluate the detector response function 
expected for an internal uniformly distributed 2.5\,keV beta, the Montecarlo was run
with few different assumptions about the mechanism causing the tail.  

Figure\,\ref{fig:beta} shows the results. For sake of comparison, in the left panel, the Montecarlo simulation was run for an external 2.5\,keV X-ray with the simple surface effect model  in (\ref{equ:diff1}) alone (light grey histogram) and with an additional short tail (darker histogram) with the parameters found for the Cl \ka\ X-ray line (see Figure\,\ref{fig:lambda} and \ref{fig:aexp}), i.e. for instance with an amplitude of about 10\%. 
The simulations for the internal uniformly distributed 2.5\,keV beta are shown in the right panel.
The simple surface model is the light grey histogram at the bottom: the
internal beta response function still presents a tail due to a small fraction
of beta interactions close to the surface, but its amplitude is about a factor
100 smaller than for the external X-ray. The darker histograms correspond to two hypotheses for the short tail origin. The black one is caused by a volume effect and therefore the tail amplitude is taken as found for the Re \ka\ escape peaks (about 70\%). 
The dark grey one is for a non-spatial effect; therefore, it is only
energy dependent and characterized by the same values found as for the external X-rays.
In both panels, the purely Gaussian peak expected at this energy is plotted for reference (solid line).

The effect of the long tail in the beta response function is negligible for an
experiment like MIBETA. As far as the short tail is concerned, the relevance
for MIBETA depends of course on which is the correct modeling. This lack of knowledge has therefore to be treated as a source of systematic uncertainty and
correctly quantified. The effect of the tails on upcoming higher statistics experiments will need a deeper analysis.

\section{Conclusions}
In this paper we have described a function which allows a satisfactory interpolation of the peaks observed in the
calibration spectra of the MIBETA experiment. 

The results of the fitting show that at least the longer of the two exponential tails towards low energy may 
be indeed understood as a result of a surface effect. 
This is confirmed by the Rhenium escape peaks which have a strongly reduced long tail.
We convincingly tested the surface effect hypothesis with a simple Montecarlo simulation. 
We used a simple phenomenological model which does not pretend to account realistically for the complex
details of electron and phonon interactions. Nevertheless the simulation proves that
a surface effect as the one described by (\ref{equ:diff1}) can explain the
observed peak shape, no matter what is the underlying physics. 

On the other hand, the shorter of
the two exponentials cannot be understood with a simple assumption. A
geometrical effect of the crystal shape may play a role, but for now we can
only conclude that we have no convincing description of this effect. 

Further
investigations with a larger number of regularly shaped crystals as of in
MARE-1 will be mandatory to clarify this situation. 

\section{Acknowledgments}
We would like to thank Prof. E. B. Norman at Lawrence Berkeley National Laboratory for providing us with the \ti\ source used for this work.

%

\end{document}